\documentclass[journal]{IEEEtran}
\usepackage{cite}
\usepackage{amsmath,amssymb,amsfonts}
\usepackage{array}
\usepackage{graphicx}
\usepackage{textcomp}
\usepackage{bm}
\usepackage[hidelinks]{hyperref}

\begin{document}

\title{On Embedding Design in Quantum Physics-Informed Neural Networks}

\author{Ban~Q.~Tran, Nahid~Binandeh~Dehaghani, Susan~Mengel, Rafal~Wisniewski, and~A.~Pedro~Aguiar%
\thanks{This work was supported in part by the Danish e-Infrastructure Consortium (DeiC) and the National Quantum Algorithm Academy (NQAA) through a Postdoctoral Scholarship under the project ``Quantum-Driven Solutions for Multi-Agent Systems and Advanced Computation''; in part by the Research Center for Systems and Technologies (SYSTEC, DOI 10.54499/UID/00147/2025) and the Associate Laboratory Advanced Production and Intelligent Systems (ARISE, DOI 10.54499/LA/P/0112/2020), funded by Funda\c{c}\~ao para a Ci\^encia e a Tecnologia, I.P./MCTES through national funds; in part by the U.S. National Science Foundation (ACCESS) Advanced Computing and Data Resource program and Texas Tech University; and by FPT University, Vietnam.}%
\thanks{B. Q. Tran and S. Mengel are with the Department of Computer Science, Texas Tech University, Lubbock, TX 79416, USA (e-mail: bantran@ttu.edu; susan.mengel@ttu.edu). B. Q. Tran is also with the Department of Computing Fundamentals, FPT University, Hanoi, Vietnam (e-mail: bantq3@fe.edu.vn).}%
\thanks{N. B. Dehaghani and R. Wisniewski are with the Department of Electronic Systems, Aalborg University, Aalborg, Denmark (e-mail: nahidbd@es.aau.dk; raf@es.aau.dk).}%
\thanks{A. P. Aguiar is with SYSTEC-ARISE, Faculty of Engineering, University of Porto, Porto, Portugal (e-mail: pedro.aguiar@fe.up.pt).}%
\thanks{Corresponding author: Ban Q. Tran (e-mail: bantran@ttu.edu).}}

\markboth{}{Tran \MakeLowercase{\textit{et al.}}: On Embedding Design in Quantum Physics-Informed Neural Networks}

\maketitle

\begin{abstract}
Quantum physics-informed neural networks (QPINNs) solve partial differential equations (PDEs) by training parameterized quantum circuits against physics-based residuals, yet the role of the embedding that maps coordinates into quantum states remains insufficiently understood. In this research, we introduce a unified embedding framework that formulates embedding as a functional transformation shaping the feature representation available to the variational circuit, and we propose two embeddings within it. The Linear Quantum Neural Network Trainable Embedding QPINN (LQNN-TE-QPINN) generates data-dependent features with an auxiliary quantum circuit and adds them to the input coordinates, instead of multiplying them as in our previous formulation. The adaptive-frequency QPINN (AdaFreq-QPINN) promotes the frequency-scaling factors of an arccos-Chebyshev encoding to trainable parameters. We evaluate both against direct and fixed analytical frequency encodings and alternative hybrid architectures on one- and two-dimensional Burgers equations. In one dimension, LQNN-TE-QPINN achieves a relative $L_2$ error of 0.0916 with 720 trainable parameters, compared with 0.4501 for direct-encoding QPINN and 0.1233 for a locally adaptive classical physics-informed neural network with approximately eleven times more parameters, while AdaFreq-QPINN attains the lowest error among the analytical embeddings with only four additional parameters. In two dimensions, LQNN-TE-QPINN attains the lowest training objective, although its solution error is not clearly separated from that of a data re-uploading QPINN. Under simulated hardware noise, LQNN-TE-QPINN retains the lowest absolute derivative error while degrading most relative to its noiseless baseline. These results demonstrate that embedding design substantially influences the approximation and optimization behavior of QPINNs.
\end{abstract}


\section{Introduction}
\label{sec:introduction}

\IEEEPARstart{P}{artial} differential equations (PDEs) play a fundamental
role in modeling a wide range of physical, biological, and engineering
systems. Classical numerical techniques, such as finite-difference and
finite-element methods, have achieved remarkable success; however,
their computational cost can become substantial for high-dimensional
or strongly nonlinear problems. Physics-informed neural networks
(PINNs) provide an alternative framework in which a neural network is
trained to satisfy the governing equations through physics-based
residuals incorporated into the loss function
\cite{raissi2019physics}. This approach enables mesh-free function
approximation and has demonstrated potential across a variety of PDE
applications, ranging from PDEs defined on complex three-dimensional
surfaces to inverse metamaterial design
\cite{fang2020surfaces,fang2019deep}.

Quantum machine learning has meanwhile explored parameterized quantum
circuits for function approximation and data-driven modeling
\cite{biamonte2017quantum,yousif2024xray}. Hybrid quantum--classical
models, which combine parameterized quantum circuits with classical
optimization, provide a practical framework for investigating such
models on current quantum devices and simulators
\cite{abughanem2024nisq,simoes2023qml,dehaghani2024hybrid}.
The integration of quantum models with physics-informed learning has
led to quantum physics-informed neural networks (QPINNs), which
incorporate parameterized quantum circuits into physics-informed
architectures
\cite{trahan2024quantum,dehaghani2025quantum,
dehaghani2026qpinnQCE,tran2026trainable,tran2026quantum}.

A fundamental component of QPINNs is the embedding, or encoding,
mechanism, which maps continuous input variables, such as spatial and
temporal coordinates, to quantum states before variational processing.
The choice of embedding determines how the input coordinates are
presented to the variational circuit and therefore shapes the feature
representation available to the overall quantum model. Consequently,
embedding design can influence the approximation properties,
trainability, and convergence behavior of QPINNs. Recent studies have
investigated structured and trainable embedding strategies in quantum
models and QPINNs
\cite{kyriienko2021solving,berger2025trainable,
dehaghani2025quantum,dehaghani2026qpinnQCE,
tran2026trainable,tran2026quantum}, with several works reporting
improved approximation performance through alternative or adaptive
feature representations. Automated searches over quantum embedding
architectures have also been investigated for supervised learning
tasks \cite{nguyen2022qes}, although a systematic comparison of
embedding families for physics-informed PDE approximation remains
limited.

Despite this progress, the relationship between embedding design and
the representation of PDE solutions in QPINNs remains insufficiently
understood. In particular, the connection between quantum embeddings
and concepts from classical function approximation, including
spectral and multi-scale representations, has received limited
attention in the QPINN setting. This question is especially relevant
for nonlinear PDEs whose solutions may contain steep gradients or
multiple characteristic scales, for which the suitability of the
induced feature representation can affect approximation quality.

A related idea appears in classical PINNs through locally adaptive
activation functions, where trainable scaling parameters modify the
effective activation scales during optimization
\cite{jagtap2020locally}. This observation motivates an analogous
relaxation at the quantum encoding stage. In analytical
frequency-based quantum embeddings, the scaling factors that determine
the frequencies introduced by the encoding~\cite{schuld2021effect} are
conventionally fixed before training. Promoting these factors to trainable parameters
provides a simple mechanism for adapting the encoding during
physics-informed optimization. This motivates the adaptive-frequency
embedding investigated in this work.

The main contributions of this work are as follows:

(i) We introduce a unified framework for embedding design in QPINNs, in which an embedding is viewed as a functional transformation that determines the feature representation presented to the variational quantum circuit. Within this framework, we establish a taxonomy spanning direct coordinate encoding, fixed analytical frequency embeddings, adaptive-frequency embeddings, and trainable quantum embeddings, which enables their systematic comparison under a common formulation.

(ii) We propose the Linear Quantum Neural Network Trainable Embedding QPINN (LQNN-TE-QPINN), a trainable quantum embedding in which an auxiliary quantum circuit generates data-dependent features that are combined linearly, by addition, with the rescaled input coordinates before encoding. This combination rule preserves each coordinate as an explicit term of the encoding angle, in contrast to the multiplicative combination of our previous work~\cite{dehaghani2026qpinnQCE}.

(iii) We propose the adaptive-frequency QPINN (AdaFreq-QPINN), an embedding that promotes the frequency-scaling parameters of an arccos-Chebyshev encoding to trainable variables, allowing the encoding frequencies to adapt jointly with the variational circuit during physics-informed optimization.

(iv) We perform a comprehensive empirical study on one- and two-dimensional Burgers equations, comparing direct encoding, the fixed analytical embeddings of the taxonomy, AdaFreq-QPINN, and LQNN-TE-QPINN, together with alternative hybrid architectures, across circuit configurations, training budgets, simulated noise channels, and finite-shot measurement settings.

(v) Through these investigations, we demonstrate that embedding design substantially influences the approximation accuracy and optimization behavior of QPINNs. The results further indicate that no single embedding is uniformly optimal, since the relative benefits of fixed, adaptive, and trainable embeddings depend on the target problem, circuit architecture, and training configuration.

The remainder of this paper is organized as follows.
Section II formulates the PDE problem and the physics-informed loss.
Section III presents the QPINN framework.
Section IV develops the generalized embedding framework and introduces
the embedding strategies considered in this work.
Section V introduces the one- and two-dimensional Burgers equations
used as benchmark problems.
Section VI presents the numerical evaluations, including circuit and
embedding sizing studies, comparisons across embedding and hybrid
architectures, and robustness analyses under simulated quantum noise
and finite measurement budgets.
Finally, Section VII summarizes the main findings and discusses
directions for future research.

\section{Problem Formulation}
\label{sec:problem_formulation}

In this work, we consider a class of time-dependent PDEs defined over a
spatial domain $\Omega \subset \mathbb{R}^d$ and a temporal domain
$t \in [0,T]$. Let $\mathbf{x}\in\Omega$ denote the spatial coordinate
vector. The governing equation is expressed in residual form as
\begin{equation*}
\mathcal{R}_{\mathrm{PDE}}[u](\mathbf{x},t)
:=
\mathcal{D}[u](\mathbf{x},t)-q(\mathbf{x},t)
=0,
\quad \!\!\!\!
(\mathbf{x},t)\in\Omega\times(0,T],
\end{equation*}
where $u(\mathbf{x},t)$ denotes the scalar solution field,
$\mathcal{D}$ is a possibly nonlinear differential operator involving
spatial and temporal derivatives, and $q(\mathbf{x},t)$ denotes a
source term.

The problem is supplemented by boundary and initial conditions,
expressed in residual form as
\begin{align*}
\mathcal{R}_{\mathrm{BC}}[u](\mathbf{x},t)
&:=
\mathcal{B}[u](\mathbf{x},t)-b(\mathbf{x},t)
=0,
\quad \!\!\!\!
(\mathbf{x},t)\in\partial\Omega\times(0,T],
\\
\mathcal{R}_{\mathrm{IC}}[u](\mathbf{x})
&:=
u(\mathbf{x},0)-u_0(\mathbf{x})
=0,
\quad \!\!\!\!
\mathbf{x}\in\Omega,
\end{align*}
where $\mathcal{B}$ denotes the boundary operator,
$b(\mathbf{x},t)$ specifies the prescribed boundary data, and
$u_0(\mathbf{x})$ specifies the initial condition.
Within the physics-informed learning framework, the solution
$u(\mathbf{x},t)$ is approximated by a parametric model
$\tilde{u}(\mathbf{x},t;\Theta)$. The parameters $\Theta$ are optimized
by minimizing the governing-equation, boundary, and initial-condition
residuals over collocation points sampled from the corresponding
domains. The resulting composite objective is
\begin{equation}
\mathcal{L}(\Theta)
=
\mathcal{L}_{\mathrm{PDE}}(\Theta)
+
\lambda_{\mathrm{BC}}\mathcal{L}_{\mathrm{BC}}(\Theta)
+
\lambda_{\mathrm{IC}}\mathcal{L}_{\mathrm{IC}}(\Theta),
\end{equation}
where
\begin{align}
\mathcal{L}_{\mathrm{PDE}}(\Theta)
&=
\frac{1}{N_{\mathrm{int}}}
\sum_{(\mathbf{x}_i,t_i)\in S_{\mathrm{int}}}
\mathcal{R}_{\mathrm{PDE}}^2[\tilde{u}]
(\mathbf{x}_i,t_i),
\\
\mathcal{L}_{\mathrm{BC}}(\Theta)
&=
\frac{1}{N_{\mathrm{bd}}}
\sum_{(\mathbf{x}_i,t_i)\in S_{\mathrm{bd}}}
\mathcal{R}_{\mathrm{BC}}^2[\tilde{u}]
(\mathbf{x}_i,t_i),
\\
\mathcal{L}_{\mathrm{IC}}(\Theta)
&=
\frac{1}{N_{\mathrm{init}}}
\sum_{\mathbf{x}_i\in S_{\mathrm{init}}}
\mathcal{R}_{\mathrm{IC}}^2[\tilde{u}]
(\mathbf{x}_i).
\end{align}
Here, $S_{\mathrm{int}}$, $S_{\mathrm{bd}}$, and $S_{\mathrm{init}}$
denote the sets of collocation points sampled from the interior,
boundary, and initial domains, respectively, while
$N_{\mathrm{int}}$, $N_{\mathrm{bd}}$, and $N_{\mathrm{init}}$
denote their cardinalities. The coefficients
$\lambda_{\mathrm{BC}},\lambda_{\mathrm{IC}}>0$
control the relative weighting of the boundary and initial-condition
terms.
The objective is therefore to construct an approximation
$\tilde{u}(\mathbf{x},t;\Theta)$ that minimizes the sampled residuals
associated with the governing equation and prescribed constraints,
without requiring labeled solution values in the interior of the
domain. In the following section, we describe how this approximation is
constructed using a hybrid quantum--classical architecture.

\section{Quantum PINN Framework}
\label{sec:qpinn_framework}

In this section, we describe the hybrid quantum--classical architecture
used to approximate the PDE solution. The model extends the
physics-informed learning framework of Section~II by employing a
parameterized quantum circuit as the function approximator. The
classical coordinates $(\mathbf{x},t)$ are first mapped to a quantum
representation through an encoding mechanism. A trainable variational
quantum circuit (VQC) then processes the encoded state, and a quantum
measurement produces the scalar approximation
$\tilde{u}(\mathbf{x},t;\Theta)$.

\subsection{Quantum State Representation}

The approximate solution is obtained from the expectation value of a
Hermitian observable $O$,
\begin{equation}
\tilde{u}(\mathbf{x},t;\Theta)
=
\langle
\psi(\mathbf{x},t;\Theta)
|
O
|
\psi(\mathbf{x},t;\Theta)
\rangle,
\end{equation}
where $|\psi(\mathbf{x},t;\Theta)\rangle$ denotes the quantum state
generated from the spatial and temporal coordinates, and $\Theta$
collects the trainable parameters of the model.
For an $n$-qubit main circuit, the quantum state is written as
\begin{equation}
|\psi(\mathbf{x},t;\Theta)\rangle
=
U_{\mathrm{var}}(\boldsymbol{\theta}_{\mathrm{var}})
\,
U_{\mathrm{enc}}
(\mathbf{x},t;\boldsymbol{\theta}_{\mathrm{emb}})
\,
|0\rangle^{\otimes n},
\end{equation}

The spatial and temporal coordinates enter the quantum
model through the input-encoding unitary $U_{\mathrm{enc}}$. For each
collocation point $(x_i,t_i)$, the coordinates are first rescaled and
then mapped to input-dependent rotation angles of the encoding gates.
Thus, different collocation points are evaluated using the same circuit
architecture and trainable parameters, but with different encoding
angles. The collocation points therefore specify where the model is
evaluated, rather than being assigned to distinct computational-basis
states. Here, $U_{\mathrm{enc}}$ denotes the input-encoding unitary and
$U_{\mathrm{var}}$ denotes the trainable variational unitary.
The parameters $\boldsymbol{\theta}_{\mathrm{emb}}$ are present only
when the embedding itself is trainable. Accordingly, the trainable
parameters may be collected as
$
\Theta
=
\left(
\boldsymbol{\theta}_{\mathrm{var}},
\boldsymbol{\theta}_{\mathrm{emb}}
\right),
$
with $\boldsymbol{\theta}_{\mathrm{emb}}$ absent for fixed embedding
strategies.
This formulation distinguishes the role of the embedding, which
determines how the classical coordinates are encoded before
variational processing, from that of the variational circuit, which
provides the trainable quantum transformation used to construct the
final input--output map.

\subsection{Variational Quantum Circuit}

The variational unitary
$U_{\mathrm{var}}(\boldsymbol{\theta}_{\mathrm{var}})$
is composed of $L$ layers,
\begin{equation}
U_{\mathrm{var}}(\boldsymbol{\theta}_{\mathrm{var}})
=
\prod_{\ell=1}^{L}
U_{\ell}(\boldsymbol{\theta}_{\ell}),
\end{equation}
where $L$ denotes the number of repeated variational layers
(i.e., the depth of the variational circuit), and
$\boldsymbol{\theta}_{\ell}$ denotes the trainable parameters of
the $\ell$-th layer. Each layer applies parameterized
$R_X$, $R_Y$, and $R_Z$ rotations to every qubit, followed by a
nearest-neighbor controlled-NOT (CNOT) entangling pattern.

Through the combined action of data encoding, variational processing,
and measurement, the resulting circuit defines a trainable nonlinear
function of the classical coordinates $(\mathbf{x},t)$.
The circuit width $n$ and depth $L$ control the size of the
variational model and influence its approximation capability,
trainability, and computational cost.

\subsection{Measurement and Output Mapping}

The scalar approximation is obtained from the expectation value of a
Hermitian observable $\mathcal{O}$. In the implementation considered
in this work, the observable is chosen as the sum of single-qubit
Pauli-$Z$ operators,
\begin{equation}
\mathcal{O}
=
\sum_{i=1}^{n}\sigma_z^{(i)},
\end{equation}
where $\sigma_z^{(i)}$ denotes the Pauli-$Z$ operator acting on the
$i$th qubit. Since the eigenvalues of $\mathcal{O}$ lie in
$[-n,n]$, the resulting expectation value satisfies
$
-n
\leq
\tilde{u}(\mathbf{x},t;\Theta)
\leq n.
$
This expectation value provides the scalar output used as the PDE
approximation.

\subsection{Physics-Informed Training}

The trainable parameters $\Theta$ are optimized by minimizing the
physics-informed objective introduced in Section~II. The temporal and
spatial derivatives of $\tilde{u}(\mathbf{x},t;\Theta)$ required by
the PDE residual are computed through the differentiable quantum model
using automatic differentiation. This enables higher-order derivatives,
such as the second-order spatial derivatives appearing in the Burgers
equations, to be incorporated directly into
$\mathcal{R}_{\mathrm{PDE}}$.

The resulting optimization is hybrid: quantum-circuit evaluations
define the parametric solution, while a classical optimizer updates
the trainable parameters according to the physics-informed objective.
In the numerical experiments, the trainable parameters are optimized
using the limited-memory Broyden--Fletcher--Goldfarb--Shanno (L-BFGS)
algorithm.

A central component of this framework is the encoding unitary
$U_{\mathrm{enc}}(\mathbf{x},t;\boldsymbol{\theta}_{\mathrm{emb}})$,
since different encoding choices provide different representations of
the classical coordinates to the variational circuit. The following
section develops a generalized formulation of this embedding stage and
introduces the fixed, adaptive-frequency, and trainable embedding
strategies investigated in the numerical study.

\section{General Embedding Framework}
\label{sec:embedding}

In QPINNs, the embedding mechanism determines how the classical
spatio-temporal coordinates are represented before being processed by
the variational quantum circuit. In this section, we introduce a unified
formulation of this mechanism and describe the embedding strategies
considered in the numerical experiments.

\subsection{Unified Embedding Formulation}

Let
$
\boldsymbol{\xi}
=
(t,x_1,\ldots,x_d)
\in\mathbb{R}^{d+1}
$
denote the complete set of spatio-temporal coordinates. Prior to
quantum encoding, the coordinates are rescaled linearly to
$[-0.95,0.95]$, which keeps them strictly inside $(-1,1)$, where
$\arccos$ and its derivative remain finite, and we denote the resulting
normalized coordinate vector
by
$
\tilde{\boldsymbol{\xi}}
=
(\tilde{t},\tilde{x}_1,\ldots,\tilde{x}_d).
$

The quantum state introduced in Section~III can then be written as
\begin{equation}
|\psi(\boldsymbol{\xi};\Theta)\rangle
=
U_{\mathrm{var}}(\boldsymbol{\theta}_{\mathrm{var}})
\,U_{\mathrm{enc}}
(\tilde{\boldsymbol{\xi}};
\boldsymbol{\theta}_{\mathrm{emb}})
\,|0\rangle^{\otimes n}.
\end{equation}
We express the encoding unitary as
\begin{equation}
U_{\mathrm{enc}}
(\tilde{\boldsymbol{\xi}};
\boldsymbol{\theta}_{\mathrm{emb}})
=
U_{\mathrm{map}}
\left(
\boldsymbol{\Gamma}
(\tilde{\boldsymbol{\xi}};
\boldsymbol{\theta}_{\mathrm{emb}})
\right),
\end{equation}
where
$
\boldsymbol{\Gamma}
(\tilde{\boldsymbol{\xi}};
\boldsymbol{\theta}_{\mathrm{emb}})
=
\mathcal{E}
(\tilde{\boldsymbol{\xi}};
\boldsymbol{\theta}_{\mathrm{emb}})
\in\mathbb{R}^{m}
$
denotes the embedded representation and $\mathcal{E}$ is the embedding
function. The complete transformation can therefore be summarized as
\begin{equation}
\boldsymbol{\xi}
\longrightarrow
\tilde{\boldsymbol{\xi}}
\xrightarrow{\;\mathcal{E}\;}
\boldsymbol{\Gamma}
\xrightarrow{\;U_{\mathrm{map}}\;}
|\psi\rangle.
\end{equation}

For fixed embeddings, no trainable embedding parameters are introduced,
whereas adaptive embeddings contain trainable parameters
$\boldsymbol{\theta}_{\mathrm{emb}}$ that are optimized jointly with
the parameters of the main variational circuit.

\subsection{Embedding and the Induced Function Space}

From a function-approximation perspective, the QPINN approximation
$\tilde{u}(\boldsymbol{\xi};\boldsymbol{\Theta})$ of the PDE solution
$u(\boldsymbol{\xi})$ can be interpreted as belonging to a family of
functions generated jointly by the embedding, variational circuit,
and measurement. Schematically, this may be expressed as
\begin{equation}
\tilde{u}(\boldsymbol{\xi};\Theta)
\approx
\sum_k c_k(\Theta)\,\phi_k(\boldsymbol{\xi}),
\end{equation}
where $\{\phi_k\}$ denote implicit basis functions associated with the
combined action of the encoding, variational processing, and
measurement. This expression is intended as a conceptual
function-approximation interpretation rather than an explicit basis
decomposition of the quantum model.

The embedding therefore influences the functional representation
available to the QPINN. This viewpoint is related to classical
approximation methods, in which the choice of polynomial, Fourier, or
other basis functions can strongly affect approximation quality
\cite{boyd2001chebyshev}. From a machine-learning perspective,
$\mathcal{E}$ may similarly be viewed as a feature map that transforms
the input coordinates into a representation used by the subsequent
model \cite{schuld2019quantum}. In particular, when the data are encoded
through Pauli-rotation gates, the model output can be written as a
truncated Fourier series in the encoded variables, whose accessible
frequencies are determined by the encoding gates and whose coefficients
are determined by the variational circuit and the measurement
\cite{schuld2021effect}. Under the arccos--Chebyshev encodings introduced
below, the encoded variable is $\arccos\tilde{\xi}_j$, and the scaling
factors of these encodings therefore determine the accessible
frequencies.
Importantly, greater flexibility of the embedding does not by itself
guarantee lower solution error. 
The realized approximation depends on how well the function space
induced by the embedding and variational circuit can represent the
spatial and temporal structure of the target solution, as well as on
the variational circuit architecture and the optimization process.

\subsection{Embedding Taxonomy}

Within this unified framework, we consider four broad classes of
embedding: (i) direct coordinate encoding, (ii) fixed analytical
frequency embeddings, (iii) adaptive-frequency encoding, and
(iv) trainable quantum embedding. These classes contain the primary
embedding strategies evaluated in the numerical study.

\subsubsection{Direct Coordinate Encoding: QPINN}

The baseline QPINN applies no additional feature transformation beyond
the coordinate rescaling introduced above. Accordingly, the embedding
is given by
\begin{equation}
\mathcal{E}_{\mathrm{ID}}(\tilde{\boldsymbol{\xi}})
=
\tilde{\boldsymbol{\xi}}.
\end{equation}
The rescaled coordinates are encoded directly as rotation angles, with
wire $i$ of the main circuit receiving the gate $R_Y(\tilde{\xi}_{j(i)})$,
where the coordinate index $j(i)$ follows the wire assignment defined in
\eqref{eq:wire_assignment}. This construction follows the
coordinate-encoding approach commonly used in differentiable quantum
models for differential equations \cite{kyriienko2021solving}.
This strategy introduces no trainable embedding parameters and serves
as the reference configuration against which the structured and
trainable embeddings are compared.

\subsubsection{Fixed Analytical Frequency Embeddings}

We next consider a family of fixed embeddings based on the
Chebyshev polynomial
$T_{\lambda}(z)=\cos(\lambda\arccos z)$, where, for a nonnegative
integer $\lambda$, $T_{\lambda}$ denotes the Chebyshev polynomial
of the first kind of degree $\lambda$. The Chebyshev polynomials
are orthogonal on $[-1,1]$ with respect to the weight
$w(z)=(1-z^2)^{-1/2}$~\cite{boyd2001chebyshev}; in the present
formulation, they are not normalized.
Let
$
\tilde{\boldsymbol{\xi}}
=
(\tilde{t},\tilde{x}_1,\ldots,\tilde{x}_d)
$
denote the rescaled spatio-temporal coordinates. Each coordinate is
encoded through rotations of the form
$
R_Y\!\left(
2\lambda_k\arccos\tilde{\xi}_j
\right),
$
following the Chebyshev-tower quantum feature-map construction
\cite{kyriienko2021solving}. For integer $\lambda_k$, the scaling
factor specifies the corresponding Chebyshev degree, or equivalently
an angular-frequency scale in the transformed variable
$\arccos(\tilde{\xi}_j)$.

To define the wire assignment uniformly across spatial dimensions, let
$D=d+1$
denote the number of spatio-temporal coordinates. For wire
$i=0,\ldots,n-1$, we define
\begin{equation}
j(i)=i\bmod D,
\qquad
k(i)=\left\lfloor\frac{i}{D}\right\rfloor,
\label{eq:wire_assignment}
\end{equation}
so that wire $i$ encodes coordinate
$\tilde{\xi}_{j(i)}$ using the scaling factor
$\lambda_{k(i)}$. Consequently, an $n$-qubit main circuit uses
\begin{equation}
P
=
\left\lceil\frac{n}{D}\right\rceil
=
\left\lceil\frac{n}{d+1}\right\rceil
\end{equation}
scaling factors.
For the one-dimensional problem, $D=2$, so the $n$ wires
alternate between $t$ and $x$, and each consecutive pair shares
one scaling factor. For the two-dimensional problem, $D=3$, so
the $n$ wires cycle through $t$, $x$, and $y$, and each consecutive
triple shares one scaling factor.
The fixed analytical embedding strategies considered in this work use
the same arccos--Chebyshev encoding and differ only in the predefined
sequence of scaling factors $\{\lambda_k\}_{k=0}^{P-1}$.

\paragraph{Tower Chebyshev (TowerCheb)}

The TowerCheb embedding uses the linearly increasing sequence
\begin{equation}
\lambda_k = k+1,
\qquad k=0,\ldots,P-1,
\end{equation}
yielding
$\{1,2,\ldots,P\}$ \cite{kyriienko2021solving}.
Through the arccos--Chebyshev encoding introduced above, these scaling
factors correspond to consecutive Chebyshev degrees and therefore
introduce a structured set of progressively increasing encoding
frequencies.
From a function-approximation perspective, this construction introduces
a predefined Chebyshev-based representation into the quantum feature
map. Chebyshev polynomial bases are widely used in approximation and
spectral methods because of their favorable approximation properties
\cite{boyd2001chebyshev}. In the present QPINN setting, however, the
Chebyshev degrees are not learned: the sequence $\{\lambda_k\}$ is fixed
before training and introduces no trainable embedding parameters.
Consequently, its effectiveness depends on how well this
predefined frequency structure represents the spatial and
temporal structure of the target solution, as well as on the
subsequent variational processing and optimization.

\paragraph{Binary}

The Binary embedding uses the dyadic sequence
\begin{equation}
\lambda_k = 2^k,
\qquad k=0,\ldots,P-1,
\end{equation}
yielding
$\{1,2,4,\ldots,2^{P-1}\}$.
This choice is motivated by exponential data-encoding strategies in
variational quantum models, where exponentially increasing encoding
scales can extend the range of frequencies introduced using a limited
number of encoding operations
\cite{shin2023exponential}.
In the present construction, the exponential scaling is applied within
the arccos--Chebyshev feature map, producing an exponentially spaced set
of encoding degrees.
Compared with the consecutive TowerCheb sequence, the Binary
construction reaches higher-order encoding degrees with the same number
of scaling factors. However, because the spacing increases
geometrically, intermediate Chebyshev degrees are not directly
introduced by the encoding. The sequence remains fixed throughout
training and introduces no trainable embedding parameters.

\paragraph{Golomb}

The Golomb embedding selects the scaling factors $\{\lambda_k\}$ from
the positive marks of a Golomb ruler, i.e., a set of integer marks whose
pairwise differences are distinct \cite{drakakis2009review}.
For example, the sequence used in the present construction begins as
$
\{1,3,7,12,\ldots\}.
$
Through the arccos--Chebyshev encoding, these scaling factors introduce
an irregularly spaced set of Chebyshev degrees. In contrast to the
consecutive TowerCheb and exponentially spaced Binary constructions,
the Golomb sequence provides a nonuniform set of encoding degrees whose
pairwise differences are non-redundant by construction. The sequence is
fixed throughout training and introduces no trainable embedding
parameters.

\paragraph{Turnpike}

We refer to the following fixed construction as the \emph{Turnpike}
embedding. It uses the triangular-number sequence
\begin{equation}
\lambda_k
=
\frac{(k+1)(k+2)}{2},
\qquad k=0,\ldots,P-1,
\end{equation}
yielding
$\{1,3,6,10,\ldots\}$.
Through the arccos--Chebyshev encoding, this construction introduces a
deterministic, nonuniform set of Chebyshev degrees whose spacing
increases progressively with $k$. Compared with TowerCheb, the sequence
reaches higher-order degrees more rapidly. It grows asymptotically less
rapidly than the Binary sequence, although for $P\le 4$, which covers
every configuration evaluated in this work, its entries $\{1,3,6,10\}$
exceed those of Binary, $\{1,2,4,8\}$, beyond the first. As with the other fixed analytical
embeddings, the sequence is predefined, remains unchanged during
training, and introduces no trainable embedding parameters.

\paragraph{Hamming}

The Hamming embedding defines its scaling factors through cumulative
Hamming weights,
\begin{equation}
\lambda_k
=
\sum_{j=1}^{k+1}\operatorname{popcount}(j),
\qquad k=0,\ldots,P-1,
\end{equation}
where $\operatorname{popcount}(j)$ denotes the number of nonzero bits
in the binary representation of $j$. This yields the sequence
$\{1,2,4,5,7,\ldots\}$.
Through the arccos--Chebyshev encoding, these scaling factors introduce
a nonuniform set of Chebyshev degrees whose growth is faster than that
of the linear TowerCheb sequence but substantially slower than that of
the exponential Binary sequence. The resulting spacing is determined
by the binary digit structure and remains fixed throughout training.
As with the other fixed analytical embeddings, Hamming introduces no
trainable embedding parameters.
For small values of $P$, the initial entries of different fixed
sequences may coincide. Consequently, different embedding labels do not
necessarily correspond to distinct encoding configurations for every
circuit width.

The five fixed analytical embeddings considered above---TowerCheb,
Binary, Golomb, Turnpike, and Hamming---differ in the distribution of
Chebyshev degrees introduced by their respective sequences
$\{\lambda_k\}$. However, they share the defining property that these
scaling factors are specified before training and remain fixed during
optimization. 
Their relative effectiveness therefore depends, among other
factors, on how well the predefined encoding structure
represents the spatial and temporal structure of the target
solution.

We relax this constraint in two stages. First, AdaFreq retains the same
analytical arccos--Chebyshev form while allowing the frequency scales
$\{\lambda_k\}$ to be learned jointly with the variational circuit.
Second, LQNN-TE replaces the predefined analytical feature
transformation with a trainable quantum embedding that generates
data-dependent features used to modify the encoded coordinates.

\subsubsection{Adaptive-Frequency Embedding: AdaFreq}

The adaptive-frequency embedding, denoted \emph{AdaFreq}, retains the
same analytical arccos--Chebyshev form as the fixed frequency
embeddings but makes the scaling factors trainable:
\begin{equation}
R_Y\!\left(
2\lambda_k\arccos\tilde{\xi}_j
\right),
\qquad
\lambda_k\in\mathbb{R}.
\end{equation}
The scaling parameters $\{\lambda_k\}$ are optimized jointly with the
variational-circuit parameters under the same physics-informed
objective.

The scaling factors are initialized as
\begin{equation}
\lambda_k=k+1,
\qquad k=0,\ldots,P-1,
\end{equation}
so that AdaFreq starts from the TowerCheb configuration and is
subsequently allowed to adapt its encoding scales during training.
Following the coordinate-grouping convention introduced above, the
number of trainable scaling factors is
$
P
=
\left\lceil\frac{n}{d+1}\right\rceil .
$
Thus, AdaFreq introduces only $P$ additional trainable parameters
relative to the corresponding fixed-embedding QPINN. In the
one-dimensional problem, each consecutive pair of wires shares one
scaling factor, whereas in the two-dimensional problem, each
consecutive triple shares one scaling factor.

At initialization, the integer values of $\lambda_k$ correspond to
consecutive Chebyshev degrees. During optimization, however,
$\lambda_k$ may take noninteger values. In this case,
$
\cos\!\left(
\lambda_k\arccos z
\right)
$
is generally no longer a polynomial in $z$. Accordingly, for
noninteger $\lambda_k$, we interpret $\lambda_k$ as a trainable
angular-frequency scale rather than as a Chebyshev polynomial degree.
AdaFreq therefore preserves the analytical cosine--arccos structure of
the fixed embeddings while allowing the encoding frequency scales to
adapt to the physics-informed objective.

This construction is conceptually related to adaptive activation
strategies in classical PINNs \cite{jagtap2020locally}, where trainable
scaling parameters modify the effective feature scales. Here, however,
the trainable scaling factors act directly on the quantum data-encoding
angles.

\subsubsection{Trainable Quantum Embedding: LQNN-TE}

The final class replaces the predefined analytical embedding with a
trainable auxiliary quantum model. Building on the trainable quantum
embedding of our previous work \cite{dehaghani2026qpinnQCE}, we propose
the Linear Quantum Neural Network Trainable Embedding (LQNN-TE), which
modifies the rule by which the auxiliary features are combined with the
input coordinates. Let
$\boldsymbol{\theta}_{\mathrm{emb}}$ denote the trainable parameters of
the embedding quantum neural network (QNN). Given the rescaled spatio-temporal coordinates
$\tilde{\boldsymbol{\xi}}$, the auxiliary circuit prepares the state
$
|\psi_{\mathrm{emb}}
(\tilde{\boldsymbol{\xi}};
\boldsymbol{\theta}_{\mathrm{emb}})\rangle,
$
from which wire-dependent expectation values are extracted. In each of
its layers, the auxiliary circuit re-uploads the rescaled coordinates
\cite{perezsalinas2020data}, encoding $\tilde{t}$ through $R_X$
rotations and each spatial coordinate through $R_Y$ rotations on the
wires assigned by the grouping $j(i)$, followed by trainable $R_X$,
$R_Y$, and $R_Z$ rotations on every qubit and a nearest-neighbor CNOT
chain.

For the $i$th main-circuit wire, the auxiliary quantum feature is
defined as
\[
b_i(\tilde{\boldsymbol{\xi}};
\boldsymbol{\theta}_{\mathrm{emb}})
=
\pi
\left\langle
\psi_{\mathrm{emb}}
(\tilde{\boldsymbol{\xi}};
\boldsymbol{\theta}_{\mathrm{emb}})
\middle|
\sigma_z^{(r(i))}
\middle|
\psi_{\mathrm{emb}}
(\tilde{\boldsymbol{\xi}};
\boldsymbol{\theta}_{\mathrm{emb}})
\right\rangle,
\]
where
$
r(i)
=
i \bmod n_{\mathrm{emb}},
$
and $n_{\mathrm{emb}}$ denotes the number of qubits in the auxiliary
embedding circuit. Since the expectation value of a Pauli-$Z$
observable lies in $[-1,1]$, the factor $\pi$ scales the auxiliary
feature to the interval $[-\pi,\pi]$.

The LQNN-TE embedding combines these data-dependent quantum features
additively with the corresponding rescaled coordinates before
encoding them into the main variational circuit. Using the
coordinate-grouping convention introduced above, the encoding angle
associated with main-circuit wire $i$ is
\begin{equation}
\gamma_i(\tilde{\boldsymbol{\xi}})
=
\tilde{\xi}_{j(i)}
+
b_i(\tilde{\boldsymbol{\xi}};
\boldsymbol{\theta}_{\mathrm{emb}}),
\label{eq:lqnn_add}
\end{equation}
where
$
j(i)=i\bmod(d+1).
$
In this formulation, $\gamma_i$ represents the rotation angle of the gate
$R_Y(\gamma_i)$ that encodes wire $i$ of the main circuit, and $j(i)$
selects the coordinate assigned to that wire.
Thus, for the one-dimensional problem, the main-circuit wires alternate
between the temporal and spatial coordinates $(t,x)$, whereas for the
two-dimensional problem they cycle through $(t,x,y)$. Each coordinate
is augmented by a data-dependent feature generated by the auxiliary
QNN. The embedding parameters
$\boldsymbol{\theta}_{\mathrm{emb}}$ and the parameters of the main
variational circuit are optimized jointly under the same
physics-informed objective.

We refer to this embedding as linear because of this combination rule.
In \eqref{eq:lqnn_add}, the auxiliary feature $b_i$ and the rescaled
coordinate $\tilde{\xi}_{j(i)}$ enter the encoding angle as a sum with
unit coefficients, so the angle is a linear combination of the two
quantities. The linearity therefore refers to how the two quantities
are combined, and $\gamma_i$ remains a nonlinear function of
$\tilde{\boldsymbol{\xi}}$ because $b_i$ is itself generated by a
quantum circuit. This rule differs from the multiplicative trainable
embedding of our previous work \cite{dehaghani2026qpinnQCE}, in which
the corresponding mapping takes the form
\begin{equation}
\gamma_i(\tilde{\boldsymbol{\xi}})
=
b_i(\tilde{\boldsymbol{\xi}};
\boldsymbol{\theta}_{\mathrm{emb}})
\,\tilde{\xi}_{j(i)}.
\label{eq:lqnn_mul}
\end{equation}
In this multiplicative form, $b_i$ acts as a data-dependent gain that
rescales the coordinate, whereas in the linear form of
\eqref{eq:lqnn_add} it acts as a data-dependent offset added to an
unchanged coordinate. The two constructions therefore define different
trainable embedding parameterizations with the same auxiliary circuit
and the same number of trainable parameters.

Unlike AdaFreq, which adapts the scaling factors within a prescribed
analytical frequency family, LQNN-TE generates data-dependent features
through a trainable auxiliary quantum circuit. It therefore provides a
different, less structurally constrained form of embedding adaptivity,
at the cost of additional trainable parameters and a larger joint
optimization problem.

\subsection{Relation to Alternative Hybrid Architectures}

In addition to the primary embedding strategies described above, the
numerical study includes several alternative hybrid architectures as
comparison models. These include a classical feed-forward
neural-network trainable embedding (FNN-TE), a quantum-assisted PINN
(QA-PINN), and a data-reuploading QPINN (Reupload-QPINN)
\cite{perezsalinas2020data}. These models
are included to distinguish the effect of trainable quantum embedding
from improvements that may arise more generally from increased model
adaptivity or alternative architectural choices. They are therefore
treated as comparison architectures rather than as members of the
primary embedding taxonomy.

\subsection{Discussion of Embedding Properties}

The considered strategies span different levels and forms of embedding
adaptivity. Direct encoding introduces no additional feature
transformation beyond coordinate rescaling, whereas TowerCheb, Binary,
Golomb, Turnpike, and Hamming introduce predefined analytical frequency
structures. AdaFreq retains the analytical cosine--arccos encoding form
while allowing the associated frequency scales to adapt during training.
In contrast, LQNN-TE employs an auxiliary trainable quantum circuit to
generate data-dependent features that are combined with the input
coordinates, providing a different and less structurally constrained
form of embedding adaptivity.

AdaFreq and LQNN-TE therefore introduce adaptivity in fundamentally
different ways. LQNN-TE uses an additional trainable quantum model and
consequently introduces a larger number of trainable parameters and a
larger optimization problem. AdaFreq, on the other hand, isolates a
more restricted form of embedding adaptivity within a fixed analytical
encoding by introducing only a small number of trainable
frequency-scaling parameters. Consequently, comparisons between
AdaFreq and the corresponding fixed analytical embeddings provide a
controlled way to examine whether adapting the encoding-frequency
scales improves approximation performance without substantially
increasing the parameter count. From this perspective, AdaFreq provides
both a lightweight adaptive embedding strategy and a mechanism for
studying the role of trainable frequency scales in QPINNs.

These constructions should not be interpreted as forming a strict
hierarchy in which greater adaptivity necessarily leads to lower
solution error. Additional embedding flexibility may improve the
realized approximation, but it can also alter the optimization
landscape and training behavior. Performance therefore depends jointly
on the embedding representation, the target PDE, the variational
circuit architecture, the optimization process, and the training
configuration.

The numerical experiments are designed to examine these trade-offs.
By comparing fixed analytical embeddings, AdaFreq, and trainable
quantum embeddings under common experimental settings, we investigate
the benefits associated with different forms of embedding adaptivity
while accounting for differences in model size and architecture. To
this end, we evaluate both the physics-informed training objective and
the resulting solution error to assess how the different embedding
strategies affect the realized approximation and optimization behavior
of QPINNs.

\section{Benchmark Nonlinear PDEs}
\label{sec:benchmark_pdes}

To evaluate the role of embedding in quantum physics-informed neural
networks, we consider time-dependent nonlinear PDEs characterized by
convective and diffusive dynamics. Such equations provide useful
benchmarks for scientific machine learning because the competition
between nonlinear transport and diffusion can generate nontrivial
solution profiles and steep spatial gradients.

A general convection--diffusion form may be expressed as
\begin{equation}
\frac{\partial u}{\partial t}
+
\mathcal{N}[u]
=
\nu\,\mathcal{D}[u],
\end{equation}
where $\mathcal{N}[\cdot]$ denotes a nonlinear convection operator,
$\mathcal{D}[\cdot]$ denotes a diffusion operator, and $\nu$ is the
diffusion coefficient. The numerical experiments in this work consider
one- and two-dimensional scalar Burgers-type equations as representative
instances of this class.

\subsection{Representative Benchmark: Burgers Equation}

The Burgers equation is a standard nonlinear convection--diffusion
model that combines nonlinear transport with viscous diffusion. Its
relatively simple mathematical form, together with its ability to
develop steep solution gradients, makes it a useful benchmark for
evaluating physics-informed function approximators.

\subsubsection{One-Dimensional Case}

For the one-dimensional benchmark, we consider
\begin{equation}
\frac{\partial u}{\partial t}
+
u\frac{\partial u}{\partial x}
=
\nu\frac{\partial^2u}{\partial x^2},
\end{equation}
defined over a spatial domain
$x\in\Omega\subset\mathbb{R}$ and a temporal interval
$t\in[0,T]$, where $\nu>0$ denotes the diffusion coefficient.
The problem is supplemented by the initial condition
\begin{equation*}
u(0,x)=-\sin(\pi x),
\end{equation*}
together with homogeneous Dirichlet boundary conditions
\begin{equation*}
u(t,x)=0,
\qquad x\in\partial\Omega.
\end{equation*}

This problem exhibits nonlinear transport moderated by viscous
diffusion and provides the primary benchmark for comparing the
different embedding strategies and circuit configurations.

\subsubsection{Two-Dimensional Case}

For the two-dimensional benchmark, we consider the scalar
Burgers-type equation
\begin{equation}
\frac{\partial u}{\partial t}
+
u\frac{\partial u}{\partial x}
+
u\frac{\partial u}{\partial y}
=
\nu
\left(
\frac{\partial^2u}{\partial x^2}
+
\frac{\partial^2u}{\partial y^2}
\right),
\end{equation}
defined over a spatial domain
$(x,y)\in\Omega\subset\mathbb{R}^2$ and a temporal interval
$t\in[0,T]$, where $\nu>0$ denotes the diffusion coefficient.
The problem is supplemented by the initial condition
\begin{equation*}
u(0,x,y)
=
-\sin(\pi x)\sin(\pi y),
\end{equation*}
together with homogeneous Dirichlet boundary conditions
\begin{equation*}
u(t,x,y)=0,
\qquad
(x,y)\in\partial\Omega.
\end{equation*}

This two-dimensional problem extends the scalar nonlinear
convection--diffusion dynamics to two spatial directions and provides
a higher-dimensional benchmark for examining the behavior of the
embedding strategies.

\begin{figure}[!t]
\centering
\includegraphics[width=\columnwidth]{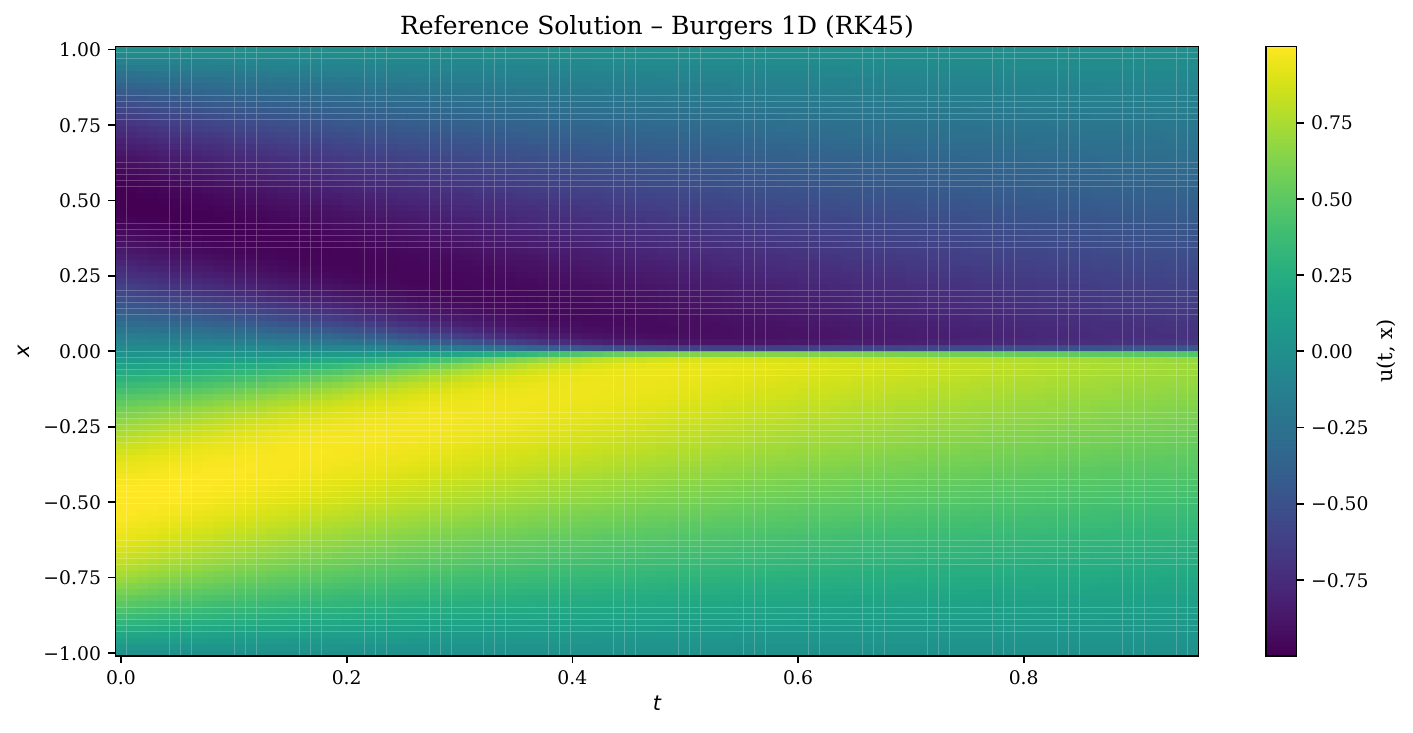}
\caption{Reference solution of the 1D Burgers equation computed with the RK45 method-of-lines scheme.}
\label{fig:ref}
\end{figure}

\begin{figure*}[t]
\centering
\includegraphics[width=\textwidth]{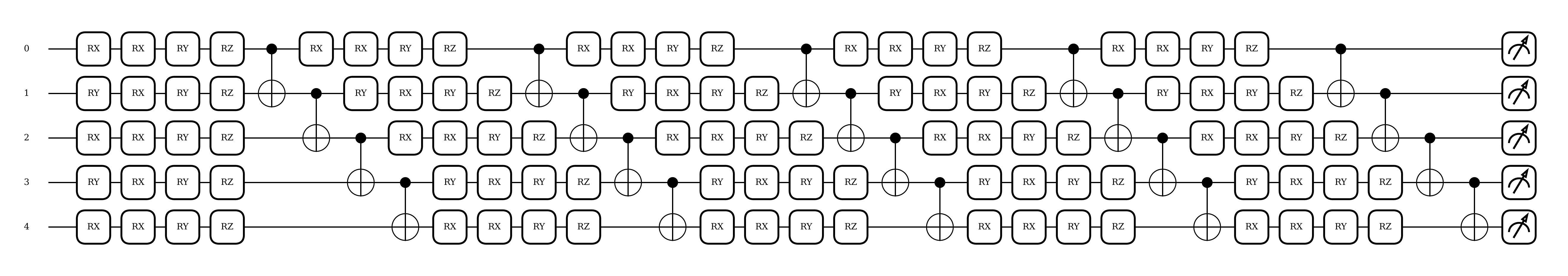}
\caption{The LQNN-TE-QPINN embedding circuit generates trainable quantum features that are supplied to the main variational quantum circuit.
}
\label{fig:qnn-emb}
\end{figure*}

\section{Numerical Results}
\label{sec:numerical_results}

In this section, we evaluate the embedding strategies and hybrid
architectures introduced in the preceding sections on the nonlinear PDE
benchmarks described in Section~V. We consider the baseline QPINN,
which directly encodes the rescaled classical coordinates into the main
variational circuit, together with the fixed analytical
embedding strategies
TowerCheb, Binary, Golomb, Turnpike, and Hamming, the adaptive-frequency
embedding AdaFreq, and the trainable quantum embedding LQNN-TE.
Additional comparisons with FNN-TE, QA-PINN, Reupload-QPINN, and
classical PINN variants are included where relevant. Throughout this
section, we refer to LQNN-TE-QPINN as the proposed model and name
AdaFreq-QPINN explicitly.

The numerical study examines several aspects of model behavior,
including the effects of main-circuit width and depth, training budget,
embedding-circuit size, and embedding choice. We further compare the
methods on the one- and two-dimensional Burgers benchmarks and assess
their behavior under simulated quantum noise and finite measurement
budgets. Because the complete model set is not evaluated under every
experimental configuration, Table~\ref{tab:coverage} summarizes the
model coverage and circuit settings used in each experiment.

\begin{table*}[!t]
\caption{Model coverage of the experiments reported in this section. Every experiment fixes one circuit configuration and one budget, so the models listed in a row are directly comparable with one another. The frequency-tower members TowerCheb, Binary, Golomb, Turnpike and Hamming differ only in the integer sequence $\{\lambda_k\}$, which is truncated to its first $P=\lceil n/(d+1) \rceil$ entries, that is $\lceil n/2 \rceil$ in one dimension and $\lceil n/3 \rceil$ in two, so several of them realize the same feature map at small qubit counts and are evaluated where they are distinguishable. All one-dimensional experiments use seeds 42, 123 and 2024; the two-dimensional experiments use seeds 42, 1004 and 2026.}
\label{tab:coverage}
\centering
\footnotesize
\renewcommand{\arraystretch}{1.25}
\newcommand{\cvL}[1]{\raggedright\arraybackslash #1}%
\begin{tabular}{|>{\raggedright\arraybackslash}p{2.6cm}|>{\raggedright\arraybackslash}p{4.4cm}|c|>{\raggedright\arraybackslash}p{7.9cm}|}
\hline
\textbf{Experiment} & \textbf{Circuit configuration} & \textbf{Epochs} & \textbf{Models evaluated} \\
\hline
Main VQC sizing (Fig.~\ref{fig:vqc_layer_qubits}) & VQC 2--8\,q $\times$ 10\,L and VQC 6\,q $\times$ 5--25\,L; embedding 6\,q $\times$ 5\,L and 4\,q $\times$ 20\,L & 100 & QPINN, TowerCheb, Turnpike, AdaFreq-QPINN, LQNN-TE-QPINN \\
\hline
Training budget (Fig.~\ref{fig:qpinn_eval}) & VQC 8\,q $\times$ 10\,L; embedding 4\,q $\times$ 5\,L & 100--500 & QPINN, TowerCheb, Turnpike, Golomb, AdaFreq-QPINN, QA-PINN, Reupload-QPINN, LQNN-TE-QPINN \\
\hline
Embedding sizing (Fig.~\ref{fig:lqnn_perf}) & VQC 6\,q $\times$ 10\,L; embedding 2--8\,q $\times$ 5\,L and 6\,q $\times$ 5--40\,L & 100 & LQNN-TE-QPINN alone, since no other model instantiates the embedding circuit \\
\hline
Embedding depth against classical baselines (Fig.~\ref{fig:sota}) & VQC 8\,q $\times$ 15\,L; embedding 4\,q $\times$ 10--50\,L & 100 & LQNN-TE-QPINN, PINN, LAAF-PINN \\
\hline
Anchor training dynamics (Figs.~\ref{fig:pinn_lqnn}, \ref{fig:qpinn_lqnn}, \ref{fig:others}) & VQC 8\,q $\times$ 15\,L; embedding 4\,q $\times$ 30\,L & 100 & PINN, LAAF-PINN, QA-PINN, QPINN, TowerCheb, Golomb, AdaFreq-QPINN, FNN-TE-QPINN, Reupload-QPINN, LQNN-TE-QPINN \\
\hline
Reconstruction from saved parameters (Fig.~\ref{fig:inference}) & VQC 8\,q $\times$ 15\,L; embedding 4\,q $\times$ 30\,L; $100 \times 100$ evaluation grid & 100 & PINN, LAAF-PINN, QA-PINN, QPINN, TowerCheb, Golomb, AdaFreq-QPINN, FNN-TE-QPINN, Reupload-QPINN, LQNN-TE-QPINN \\
\hline
Two-dimensional Burgers (Figs.~\ref{fig:2d-training}, \ref{fig:2d-embedding}; Table~\ref{tab:2d200}) & VQC 5\,q $\times$ 15\,L; embedding 4\,q $\times$ 30\,L & 100 / 200 & QPINN, Turnpike, Hamming, AdaFreq-QPINN, QA-PINN, Reupload-QPINN, LQNN-TE-QPINN \\
\hline
Noise channels and finite shots (Figs.~\ref{fig:noise}, \ref{fig:finite-shot}) & VQC 8\,q $\times$ 15\,L; embedding 4\,q $\times$ 30\,L & 100 & QPINN, AdaFreq-QPINN, LQNN-TE-QPINN for the channels, LQNN-TE-QPINN for the shot budget \\
\hline
\end{tabular}
\end{table*}

\subsection{Environment}

\begin{figure*}[t]
\centering
\includegraphics[width=0.8\textwidth]{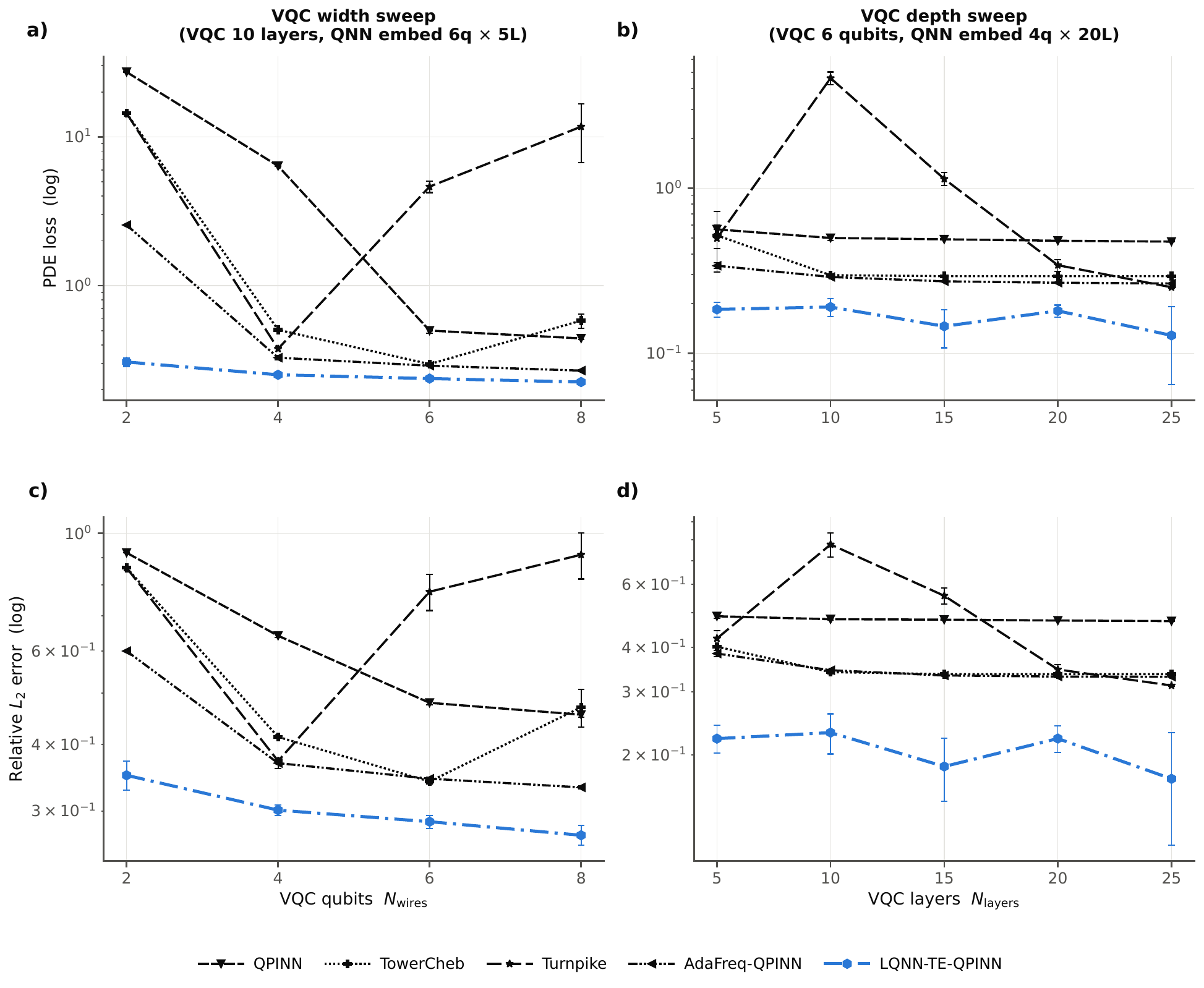}
\caption{Performance exploration of the main VQC in QPINN for solving the 1D Burgers equation. (a) and (c) VQC sweep with 10 layers across varying qubit counts from 2 to 8. (b) and (d) VQC sweep with 6 qubits across varying layer counts from 5 to 25. The top row shows the solver performance in terms of PDE loss. The bottom row presents the relative $L_2$ error of the predicted solution with respect to the RK45 reference solution. The two sweeps pin different embedding circuits, 6 qubits by 5 layers for the left column and 4 qubits by 20 layers for the right column, so points in the two columns are not directly comparable and the LQNN-TE-QPINN values of the two columns must not be pooled.}
\label{fig:vqc_layer_qubits}
\end{figure*}

All experiments in this study were carried out on classical hardware running quantum simulators, a choice dictated by the present noisy intermediate-scale quantum (NISQ) era, in which cost-effective, fault-tolerant quantum processors are not yet available for model training \cite{abughanem2024nisq}. We performed the simulations on the REmotely-managed Power Aware Computing Systems and Services (REPACSS) supercomputer hosted at Texas Tech University \cite{repacss}, drawing on its graphics processing unit (GPU) compute nodes, each provisioned with four NVIDIA H100 NVL accelerators offering 94 GB of high-bandwidth memory per device. Additionally, we leveraged the computing resources of the Bridges-2 GPU cluster at the Pittsburgh Supercomputing Center, equipped with 192 NVIDIA V100-32GB SXM2 GPUs and 80 H100-80GB SXM5 GPUs \cite{psu}.
For the circuit sizes considered in this study, classical simulation
was feasible for most configurations, although memory limitations arose
for some wider or deeper models due to the combined cost of
state-vector simulation and higher-order automatic differentiation.
On the software side, the classical neural network components and the hybrid optimizer were implemented in Python using the PyTorch deep learning framework \cite{pytorch}, whereas the quantum neural network architecture was built with the PennyLane library, which we selected for its ease of use, reliability, and broad hardware support during the post-training inference stage \cite{pennylane}.

To measure the accuracy of the predicted solutions from all embedding methods, we use the classical explicit Runge--Kutta method of order 5(4) (RK45) to compute the reference solution of the Burgers equation over the time domain $[0, 0.95]$ and the spatial domain $[-1, 1]$, at a viscosity of $\nu = 0.01/\pi \approx 3.18 \times 10^{-3}$ (Figure~\ref{fig:ref}). At this viscosity a central advection scheme is unstable, so the reference is obtained by the method of lines with first-order upwind advection on the collocation grid, and the solution develops a steep internal front at $x = 0$ near $t \approx 1/\pi \approx 0.32$. Unless stated otherwise, $\pm$ denotes one standard deviation computed
as the population standard deviation over the three seeds or repeats;
the sizing sweeps of Figures~\ref{fig:vqc_layer_qubits},
\ref{fig:lqnn_perf} and \ref{fig:sota} report the sample standard
deviation. All QPINN models use a hardware-efficient quantum circuit ansatz to enable the main VQC to learn its parameters. 
The sweeps reported below cover main-circuit widths from 2 to 8 qubits and depths from 5 to 25 layers, so 2 qubits and 5 layers are the smallest configurations tested rather than a claimed lower bound on what the method needs.
Wider circuits were not reached because state-vector simulation under reverse-mode differentiation is memory-bound, so larger configurations become computationally demanding within our classical simulation setup.
Over the tested range, increasing the VQC layer depth does not substantially reduce the solution error for the Burgers equation, whereas the circuit width has a stronger effect.
Among the configurations tested on our H100 GPU setup, the selected configuration for the main VQC is 8 qubits and 15 layers. For the trainable embedding circuit of LQNN-TE-QPINN, the best-performing configuration among those tested, detailed in subsequent sections, is 4 qubits and 30 layers. Figure~\ref{fig:qnn-emb} illustrates the LQNN-TE-QPINN embedding circuit at 5 qubits and 5 layers, a size chosen for readability rather than one used in the experiments.

\subsection{One-Dimensional Training Evaluation}
In the following section, we conduct a comprehensive evaluation of the LQNN-TE-QPINN model's performance in solving the 1D Burgers PDE and elaborate on the configuration selected for this problem. We evaluate the proposed model under simulated noise and finite measurement budgets separately in Section~VI-D.

\subsubsection{Sizing the Main VQC: Width versus Depth}
Figure~\ref{fig:vqc_layer_qubits}(a) and (c) sweep $N_{\mathrm{wires}}$ of the main variational quantum circuit (VQC) from 2 to 8 at a fixed depth of 10 layers, whereas (b) and (d) sweep $N_{\mathrm{layers}}$ from 5 to 25 at a fixed width of 6 qubits, and we train every configuration for 100 epochs with the L-BFGS optimizer over three random seeds, drawing error bars of one standard deviation. 
Within the tested configurations, width has a stronger effect on solution error than depth:
the baseline QPINN reduces its relative $L_2$ error from $0.9193$ at 2 qubits to $0.4554$ at 8 qubits, a $50.5\%$ reduction, while gaining only $3.1\%$ between 5 and 25 layers, and AdaFreq-QPINN and TowerCheb corroborate this asymmetry with depth gains of $13.9\%$ and $16.2\%$ over the same range. We qualify the width reduction, since the 2-qubit baseline carries a PDE loss of $27.196$, so part of that $50.5\%$ records the circuit becoming trainable at all. Widening the circuit is nevertheless not a uniformly beneficial choice, since Turnpike performs best at 4 qubits ($0.3726$) and degrades by a factor of $2.4$ to $0.9114$ at 8 qubits, TowerCheb performs best at 6 qubits ($0.3408$) and worsens by $38\%$ to $0.4701$ at 8, while only the adaptive encodings (LQNN-TE-QPINN and AdaFreq-QPINN) and the direct baseline improve monotonically to 8 qubits.
A possible explanation is that the additional predefined encoding
frequencies interact less favorably with the target solution and
subsequent variational optimization at these wider configurations.

\begin{figure}[!t]
\centering
\includegraphics[width=\columnwidth]{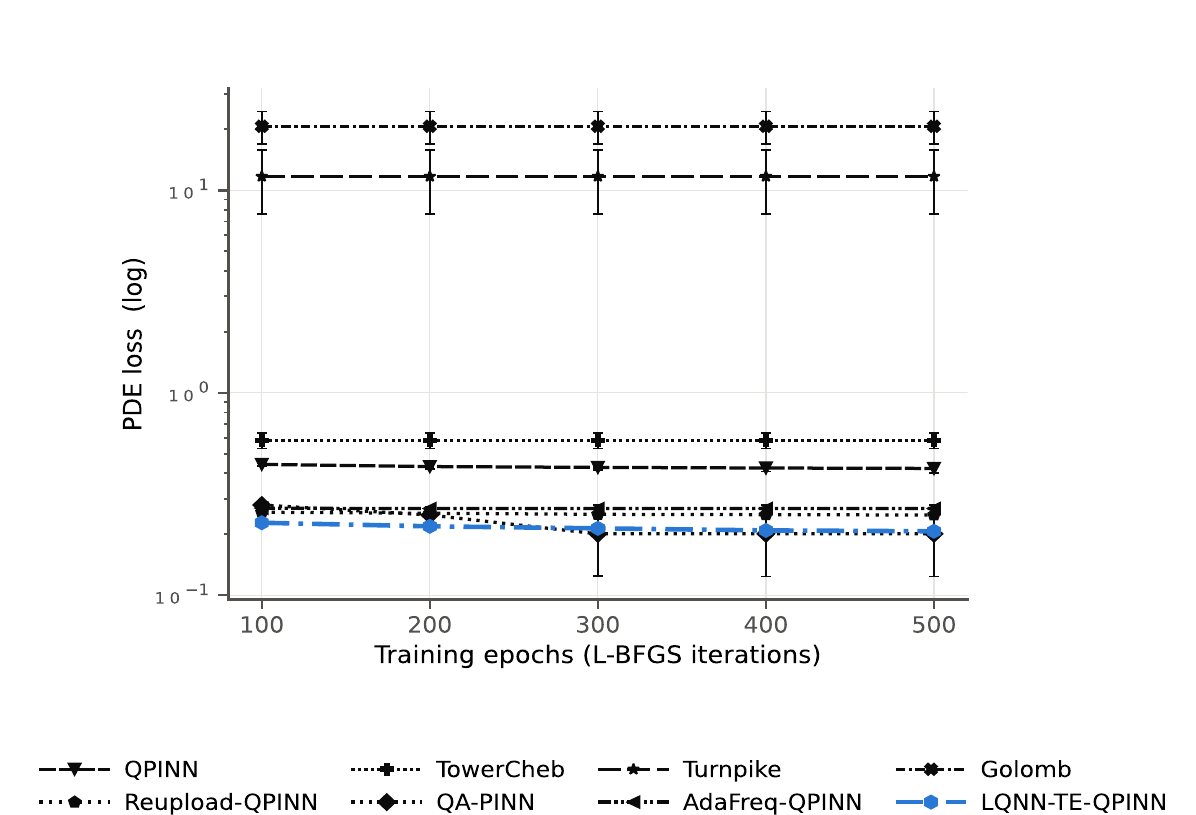}
\caption{Performance exploration of quantum solvers across training 
budgets 
ranging from 100 to 500 epochs.}
\label{fig:qpinn_eval}
\end{figure}

The proposed LQNN-TE-QPINN traces the lowest curve in all four panels and is also the flattest in width, improving by $22.9\%$ against the $50.5\%$ recorded by the baseline. At 2 qubits and 150 trainable parameters it reaches a relative $L_2$ error of $0.3501$, which falls $23\%$ below the $0.4554$ that the baseline QPINN requires 8 qubits and 240 parameters to reach. The trainable embedding therefore appears to carry part of the approximation burden that the baseline must obtain from solver size, which is consistent with our broader observation that embedding design can substantially influence approximation performance. We report one caveat directly: the depth response of the LQNN-TE-QPINN is non-monotone, its seed standard deviation at 25 layers reaches $0.0598$, roughly $35\%$ of the mean, and the per-seed winners disagree, with seeds 42 and 2024 favoring 25 layers and seed 123 favoring 15, so we cannot name a best depth for this model. We therefore adopt 8 qubits and a depth of 10 to 15 layers for the main VQC, since past 15 layers the non-LQNN curves vary by at most $0.005$ in relative $L_2$ error. For the fixed analytical towers, 4 and 6 qubits are the best-performing tested widths, and we nevertheless evaluate them at 8 qubits in the anchor comparisons below in order to test every embedding at the largest qubit count that all models support in our simulation setup.

\begin{figure*}[t]
\centering
\includegraphics[width=0.8\textwidth]{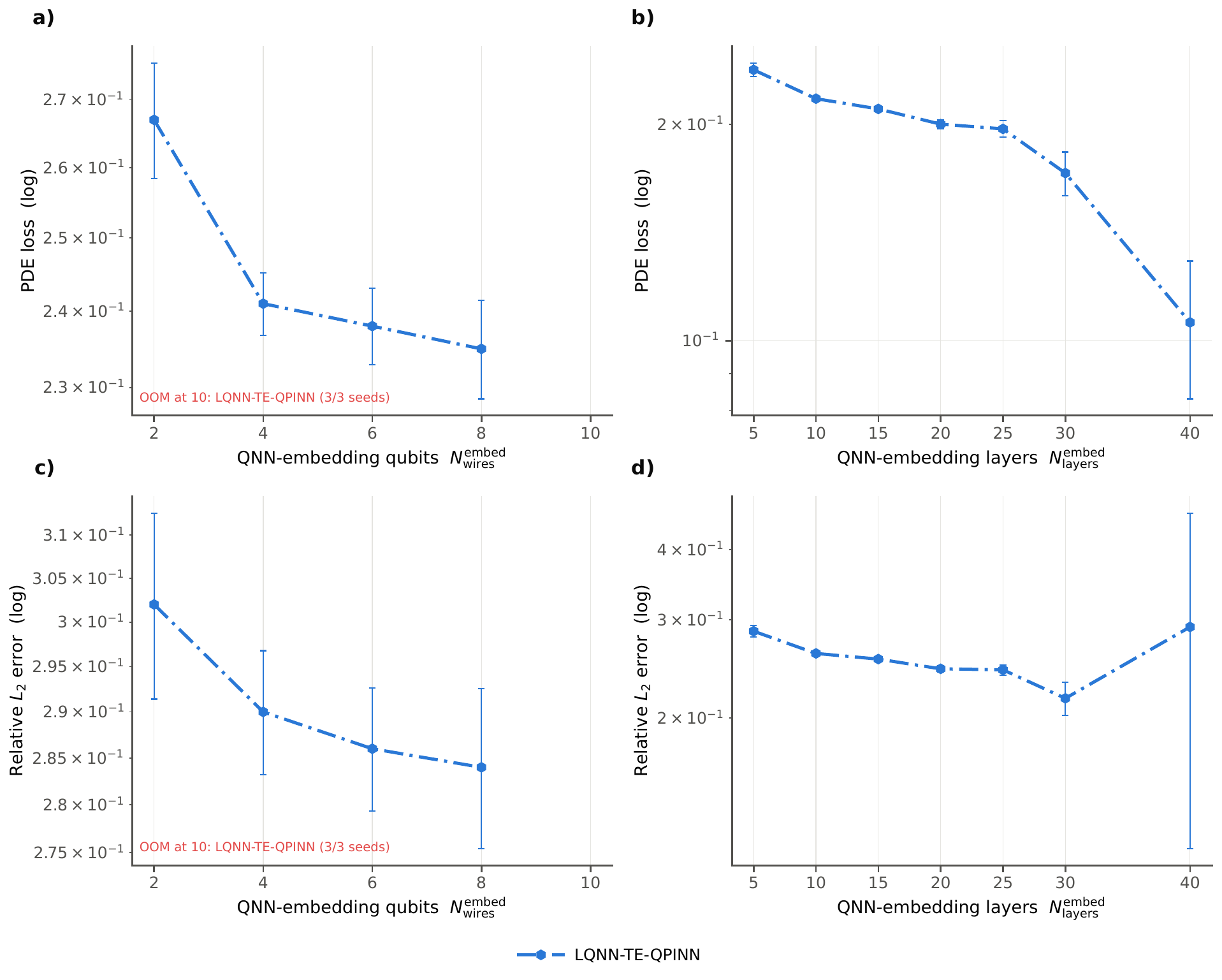}
\caption{Performance evaluation of the LQNN-TE-QPINN solver across various qubit and layer configurations, using a fixed main VQC setup of 6 qubits and 10 layers, trained for 100 epochs. (a) and (c) QNN architecture with a fixed depth of 5 layers and varying qubit counts from 2 to 8. (b) and (d) QNN architecture with a fixed width of 6 qubits and varying layer counts from 5 to 40.}
\label{fig:lqnn_perf}
\end{figure*}

\subsubsection{Sizing the Trainable Embedding Circuit}
We first examine whether increasing the training budget changes the
observed model ordering. Figure~\ref{fig:qpinn_eval} illustrates the
PDE loss of each quantum solver across L-BFGS training budgets ranging
from 100 to 500 epochs.
LQNN-TE-QPINN improves its relative $L_2$ error from $0.273$ to $0.247$, a gain of only $9.5\%$ for five times the cost, the baseline QPINN moves from $0.455$ to $0.444$, and TowerCheb, Turnpike, Golomb and AdaFreq-QPINN remain flat to three decimals. 

These results indicate that the model ordering remains unchanged
over the tested range of 100--500 L-BFGS epochs. We therefore
adopt 100 epochs for the subsequent experiments and interpret
the following sweeps as differences that are not removed simply by
increasing the optimization budget over the range considered here.
Panels (a) and (c) of Figure~\ref{fig:lqnn_perf} vary $N_{\mathrm{wires}}^{\mathrm{embed}}$ from 2 to 8 at five embedding layers, the main VQC pinned at 6 qubits and 10 layers. The PDE loss falls from $0.267 \pm 0.0086$ at two qubits to $0.241 \pm 0.0042$ at four, a $10\%$ reduction worth about $3.7\sigma$ of the seed spread, after which the axis goes inert: from four to eight qubits the loss moves $0.006$ in total, roughly $1.1\sigma$. The per-seed records are decisive: seed 42 ranks eight qubits first ($0.227$), seed 123 places six ahead of eight ($0.234$ against $0.243$), and seed 2024 four ahead of six ($0.244$ against $0.245$). The advantage of eight qubits survives only under averaging, so we adopt four embedding qubits, which captures the whole available gain with 240 parameters against 300.

\begin{figure*}[t]
\centering
\includegraphics[width=\textwidth]{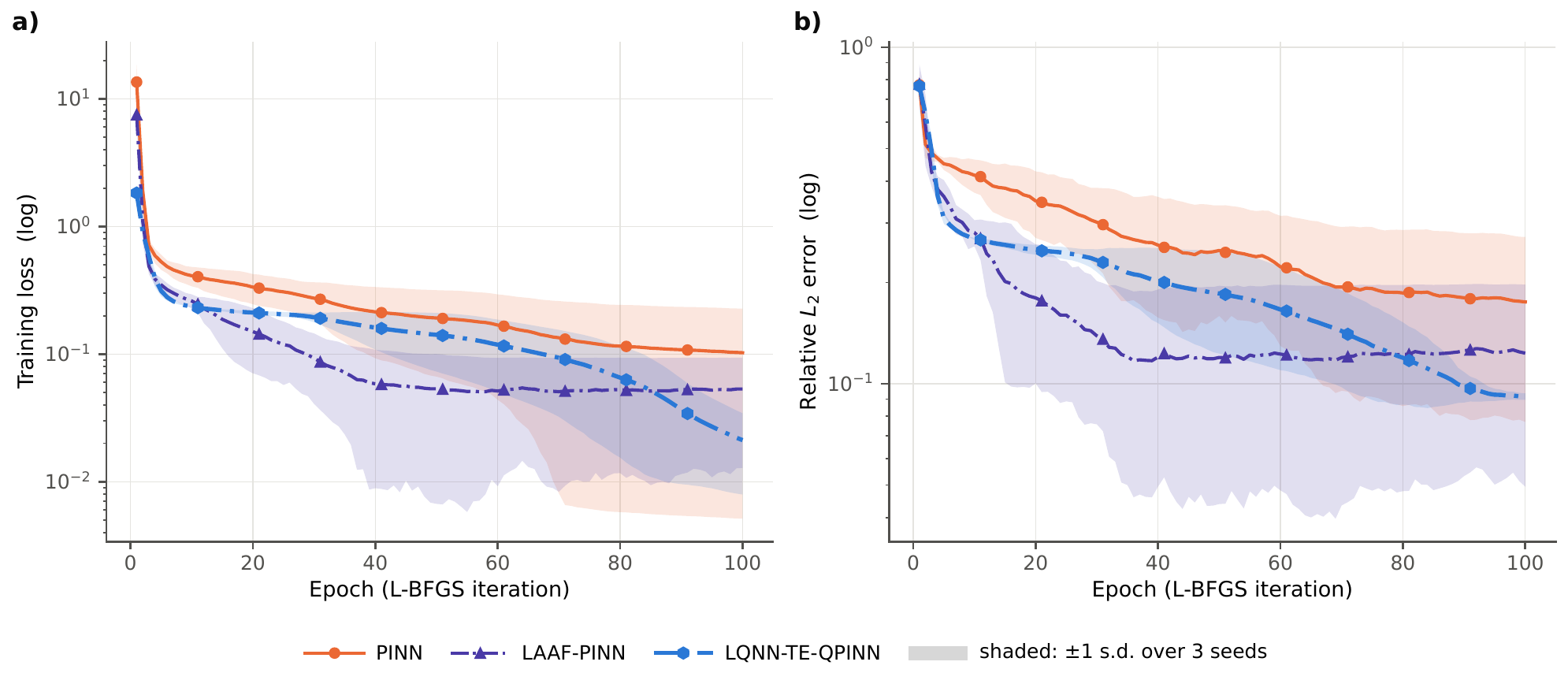}
\caption{Performance comparison of the LQNN-TE-QPINN solver against classical PINN and the locally adaptive LAAF-PINN variant. (a) PDE loss progression over 100 training epochs. (b) Solution error evaluated against the RK45 reference solution using the relative $L_2$ error metric. All experiments were conducted with 3 random seeds, and the averaged results are reported. }
\label{fig:pinn_lqnn}
\end{figure*}

\begin{figure}[!t]
\centering
\includegraphics[width=0.4\textwidth]{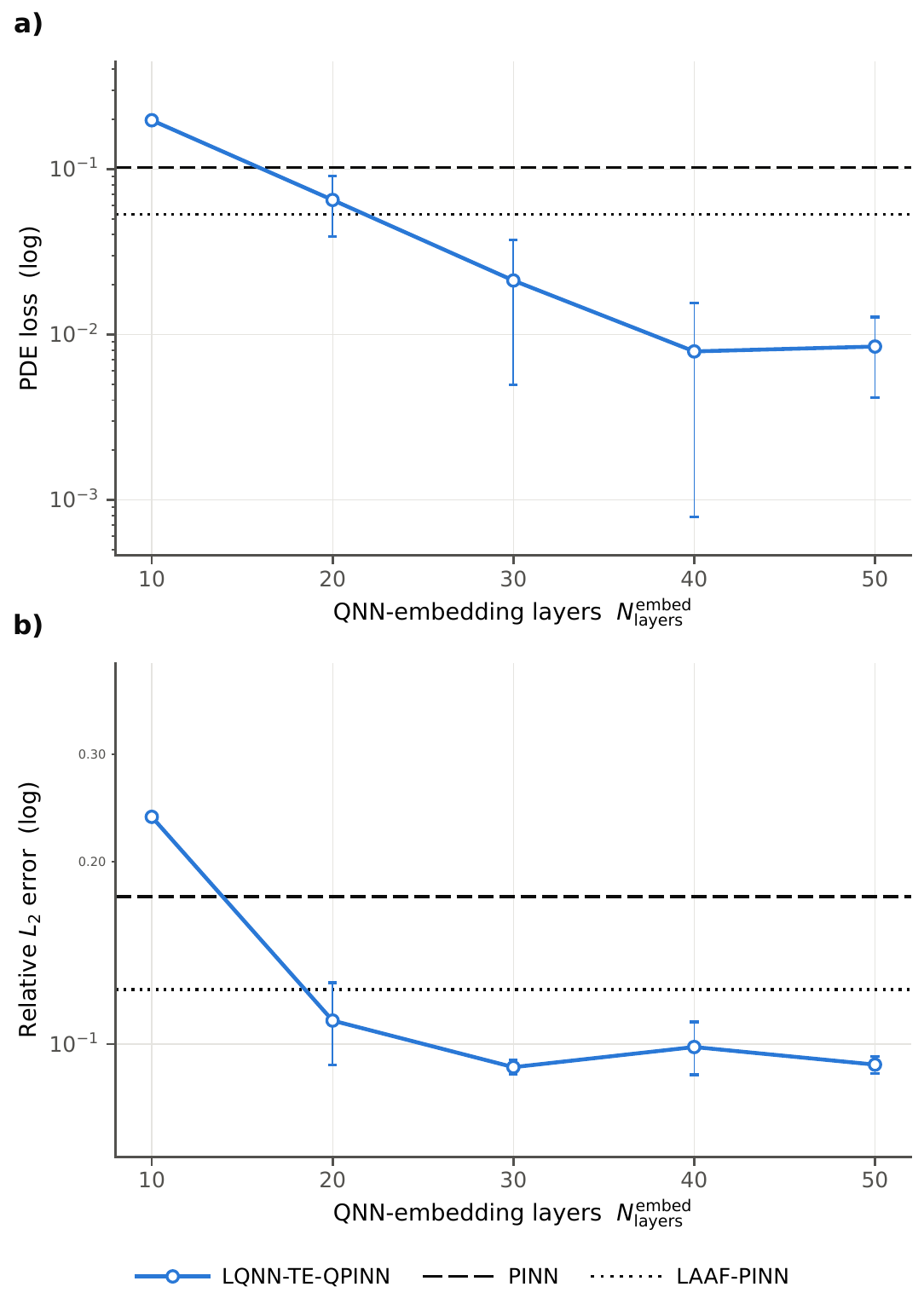}
\caption{Embedding depth at which LQNN-TE-QPINN overtakes the classical solvers during the initial 100 training epochs. The QNN architecture is held fixed at 4 qubits while varying the layer count from 10 to 50. (a) PDE loss progression. (b) Relative $L_2$ error evaluation.}
\label{fig:sota}
\end{figure}

Panels (b) and (d) vary $N_{\mathrm{layers}}^{\mathrm{embed}}$ from 5 to 40 at six embedding qubits, where the trend is genuine. The PDE loss declines from $0.238 \pm 0.0051$ to $0.171 \pm 0.0119$ at thirty layers, which wins on all three seeds and both metrics, the closing step from 25 to 30 contributing $-0.026$, about $2.8\sigma$. At forty layers the two metrics point in opposite directions: the loss reaches $0.106$, the lowest value of that sweep, while the relative $L_2$ error regresses to $0.291$ and its standard deviation explodes from $0.015$ to $0.174$. Seed 42 drives the objective to $0.0737$, the lowest single loss of that sweep, while returning $0.536$, its worst relative $L_2$ error, whereas seeds 123 and 2024 reach the two best values of that sweep, $0.170$ and $0.165$; the seed-to-seed ratio of that error, $1.02$ to $1.05$ up to 25 layers and $1.18$ at 30, reaches $3.25$ at 40. What the data establish at this depth is that the physics-informed training loss and the solution error rank the seeds differently, since the seed reaching the lowest objective returns the worst relative $L_2$ error of the sweep. A lower physics-informed loss therefore does not correspond to a lower solution error at forty layers, and we report this divergence rather than attributing it to a particular mechanism. Figure~\ref{fig:sota} then measures the joint setting directly at four embedding qubits with the main VQC at its 8-qubit, 15-layer anchor, a point neither sweep of Figure~\ref{fig:lqnn_perf} contains. There the relative $L_2$ error attains its minimum of $0.0916$ at thirty layers, which deeper circuits do not beat ($0.0989$ at 40, $0.0925$ at 50) even as the PDE loss keeps falling from $0.0212$ to $0.0079$, reproducing the divergence. Thirty layers is also the cheapest setting reaching that accuracy, 720 parameters against 840 and 960, and it carries the tightest seed spread of any depth beyond ten, $0.0025$ against $0.0170$ at twenty and $0.0099$ at forty. We consequently adopt 4 qubits by 30 layers.

We report two sweep points that exhaust the memory of the simulator. LQNN-TE-QPINN fails at $N_{\mathrm{wires}}^{\mathrm{embed}} = 10$ on all three seeds, as annotated in panels (a) and (c), while the models that do not instantiate the embedding circuit train normally there; Reupload-QPINN fails at a main VQC width of nine qubits, where all others train. This ceiling may be attributed to our differentiation requirements: the residual needs second-order derivatives of the circuit output with respect to its inputs, forcing state-vector simulation with a backpropagation-capable method under the PyTorch interface. Reverse-mode differentiation retains the tape's intermediate states, so the footprint grows as $O(2^{n})$ per state and multiplies by circuit depth and collocation batch. LQNN-TE-QPINN, the only model instantiating a second circuit, therefore reaches that ceiling on the embedding-width axis, whereas Reupload-QPINN re-encodes the coordinates inside the solver circuit without intermediate readout and carries the deepest single tape. Both failures therefore reflect a limit of our classical simulation and differentiation setup rather than a property of the models themselves. The two sweeps of Figure~\ref{fig:lqnn_perf} also form an L crossing only at six qubits and five layers, so neither contains the adopted configuration, and any pairing of the two one-dimensional optima remains an extrapolation. Figure~\ref{fig:sota} mitigates this by supplying the direct measurement at four qubits, and a grid of qubits in $\{4,6,8\}$ by layers in $\{10,20,30\}$ would settle the interaction between the two axes.

\begin{figure*}[t]
\centering
\includegraphics[width=\textwidth]{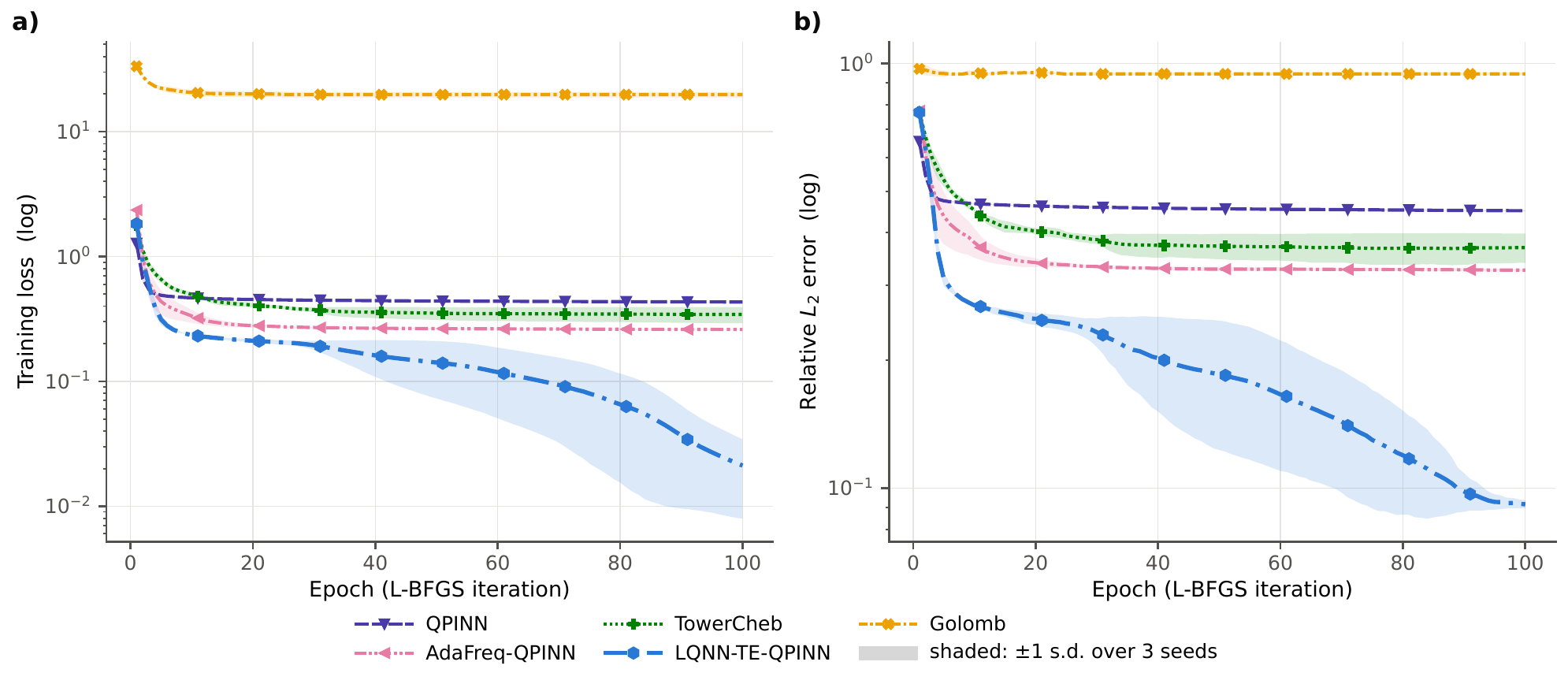}
\caption{Performance evaluation of the LQNN-TE-QPINN solver on the Burgers PDE compared to alternative embedding methods utilizing analytical functions instead of a quantum model. (a) Training dynamics evaluated by PDE loss. (b) Solution accuracy measured via relative $L_2$ error. All experiments were conducted across multiple random seeds using the L-BFGS optimizer. }
\label{fig:qpinn_lqnn}
\end{figure*}
\subsubsection{Comparison against Classical PINN Baselines}
Figure~\ref{fig:pinn_lqnn} illustrates the mean training trajectories of PINN, the locally adaptive activation function PINN (LAAF-PINN) \cite{jagtap2020locally} and LQNN-TE-QPINN over 100 L-BFGS epochs, with a $\pm 1$ standard deviation band across seeds 42, 123 and 2024, at our adopted configuration of a main VQC of 8 qubits and 15 layers together with a trainable embedding of 4 qubits and 30 layers. We track the epoch from which the mean curve of LQNN-TE-QPINN remains permanently below each baseline: epoch 4 against PINN, and only epoch 79 against LAAF-PINN, the strongest classical baseline in this comparison. The late crossing is the more informative of the two. Through the first half of the budget LAAF-PINN holds a clear lead, reaching a relative $L_2$ error of $0.1800$ at epoch 20 and $0.1201$ at epoch 40 against $0.2504$ and $0.2023$ for our model, and it then stops improving, moving only from $0.1201$ to $0.1233$ between epoch 40 and epoch 100. LQNN-TE-QPINN descends steadily over the same interval, from $0.2023$ to $0.0916$, and the crossover therefore stems from the classical baseline saturating while the trainable quantum embedding is still learning. That descent has not ended at the edge of the budget, since the model first comes within $5\%$ of its final relative $L_2$ error at epoch 77 on average and at epoch 95 on seed 42.

The means alone would understate this result. At epoch 100 LQNN-TE-QPINN reaches $0.0916 \pm 0.0021$ against $0.1233 \pm 0.0739$ for LAAF-PINN and $0.1751 \pm 0.0981$ for PINN. The means differ by a few hundredths, whereas the standard deviation of our model is $47\times$ smaller than that of PINN and $35\times$ smaller than that of LAAF-PINN. The per-seed records make the difference concrete: LAAF-PINN records $0.0629$, $0.0797$ and $0.2273$, so it is the more accurate model on two seeds out of three and considerably worse on the third, while LQNN-TE-QPINN records $0.0933$, $0.0887$ and $0.0928$ and lands within $0.005$ of the same answer every time. This collapse of the spread is not a final-epoch artifact, since averaged over epochs 20 to 100 the seed spread is $0.035$ for LQNN-TE-QPINN against $0.073$ for LAAF-PINN and $0.095$ for PINN. We state the claim as a conjunction, because either half alone is satisfied by a weaker model: at this configuration LQNN-TE-QPINN attains both the lowest final error and a seed spread that collapses at convergence, and it does so with 720 trainable parameters against 8,051 for LAAF-PINN and 7,851 for PINN, roughly $11\times$ fewer.

Figure~\ref{fig:sota} quantifies the embedding depth at which this advantage appears, sweeping the trainable embedding from 10 to 50 layers at 4 qubits with the main VQC held at the anchor and the classical baselines drawn as horizontal lines at their 100-epoch values. At 10 layers and 480 parameters the quantum model trails both classical solvers, at $0.2369$ against $0.175$ for PINN and $0.123$ for LAAF-PINN. At 20 layers it passes both at $0.1093$ with 600 parameters, roughly a thirteenth of the classical budget, and 30 layers gives the best measured accuracy at $0.0916$, while 40 and 50 layers add parameters without adding accuracy ($0.0989$ and $0.0925$). The PDE loss crosses at different depths, passing PINN at 20 layers ($0.0649$ against $0.102$) and LAAF-PINN only at 30 layers ($0.0212$ against $0.0533$), and since the loss and the relative $L_2$ error disagree about the winner on all three seeds, we let the relative $L_2$ error carry the ranking. We acknowledge that this advantage holds at a sufficient embedding depth and is not unconditional. At our earlier and smaller anchor, a main VQC of 6 qubits and 10 layers with a 6-qubit, 5-layer embedding, LQNN-TE-QPINN reached only $0.286$ and sat behind both LAAF-PINN ($0.123$) and PINN ($0.175$). A plausible explanation is that a shallow embedding cannot supply a feature map rich enough for the main circuit to exploit. We mitigate this dependence by reporting the threshold measured in
Figure~\ref{fig:sota}, namely 20 embedding layers at 4 qubits as the
smallest tested depth at which our model overtakes both classical
baselines, and by attaching the configuration to every reliability
claim, since seed sensitivity is a property of a model at a given
configuration rather than a fixed trait. A longer training budget
would help determine whether the performance observed at 100
epochs remains stable with further optimization.
\begin{figure*}[t]
\centering
\includegraphics[width=\textwidth]{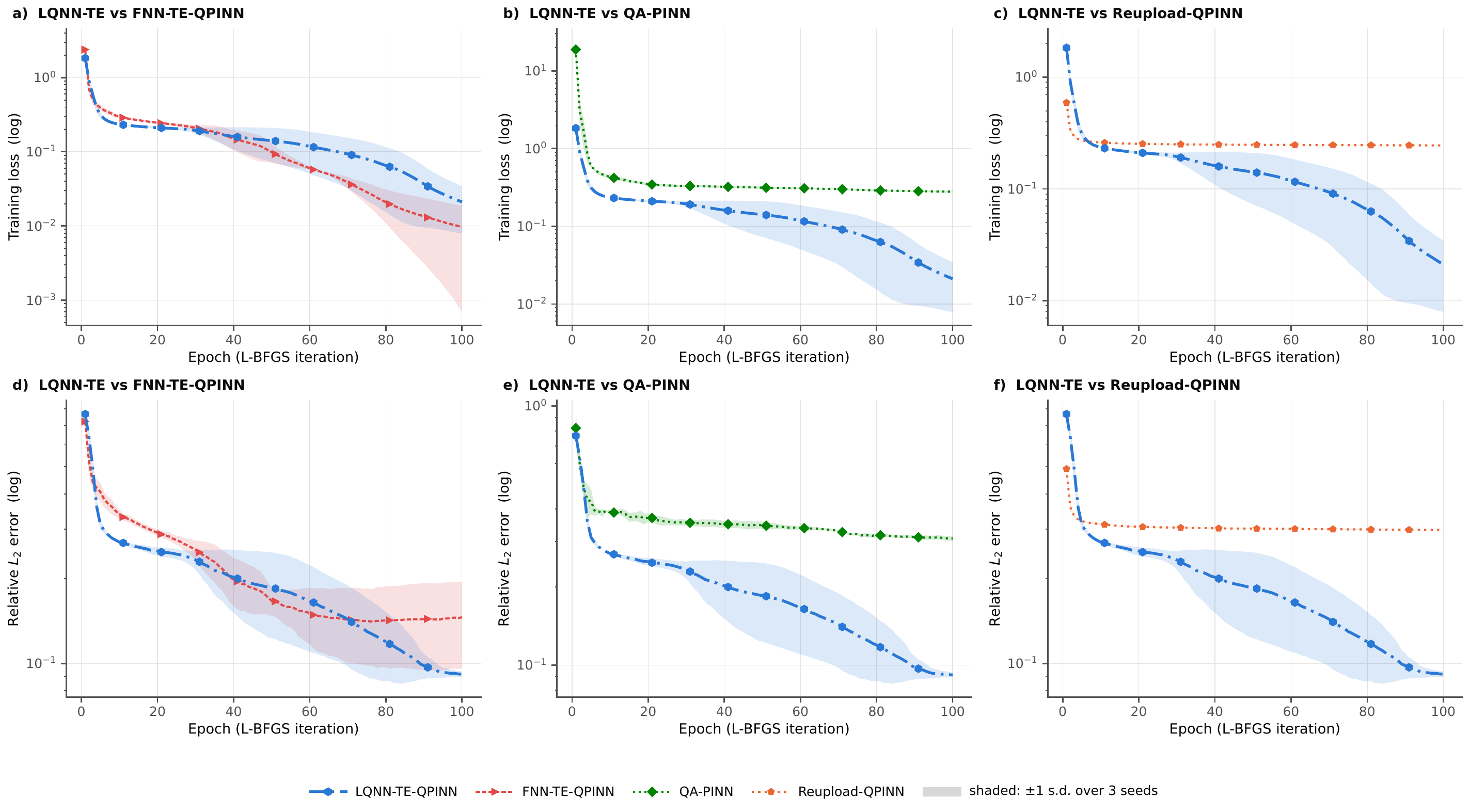}
\caption{Performance evaluation of the LQNN-TE-QPINN solver against alternative solver architectures over 100 L-BFGS epochs, with mean curves and $\pm 1$ standard deviation bands over seeds 42, 123 and 2024. (a) and (d) a classical neural network generating the quantum feature map (FNN-TE-QPINN); (b) and (e) a PINN with one hidden layer replaced by a quantum circuit (QA-PINN); (c) and (f) data re-uploading (Reupload-QPINN). The top row reports the training loss and the bottom row the relative $L_2$ error.}
\label{fig:others}
\end{figure*}

\subsubsection{Comparison against Analytical Embedding Strategies}
Figure~\ref{fig:qpinn_lqnn} compares five embedding strategies on the 1D Burgers equation, with every series running on the identical main VQC of 8 qubits and 15 layers under an identical budget of 100 L-BFGS epochs over three seeds (42, 123, 2024), so the embedding is the only variable. At epoch 100 the ranking by relative $L_2$ error is unambiguous: LQNN-TE-QPINN reaches $0.0916 \pm 0.0021$, AdaFreq-QPINN $0.3262 \pm 0.0020$, TowerCheb $0.3684 \pm 0.0293$, the identity-encoding baseline QPINN $0.4501 \pm 0.0013$, and Golomb $0.9452 \pm 0.0074$. The trainable quantum embedding is therefore $3.6\times$ more accurate than AdaFreq-QPINN, the strongest analytical embedding here, $4.0\times$ better than the 
TowerCheb embedding and $4.9\times$ better than the identity baseline, at $2\times$ the parameters of any of them, 720 against roughly 360. We report the Golomb failure directly: it attains its final value at epoch 1 on all three seeds and never moves again, at a training loss of $19.8188$ that sits three orders of magnitude above every other series. Its error is not exactly $1.0$, so the output is not identically zero and the model runs while learning essentially nothing at this width. It trained normally at 4 qubits, reaching $0.373$, so
we interpret this as a configuration-dependent width degradation of the Golomb embedding at eight qubits rather than as evidence of a broken run. We evaluate the fixed towers at this width deliberately, since 8 qubits is the largest main circuit that every model in this comparison trains in our simulation setup.
Because its loss is disproportionate to its error, we rank it by relative $L_2$ error alone.

The convergence trajectories provide additional context for
this ranking, and the epoch at which each model first comes within $5\%$ of its own final relative $L_2$ error makes it explicit. Golomb arrives at epoch 1, the identity baseline at epoch 7, AdaFreq-QPINN at epoch 16 and TowerCheb at epoch 27, whereas LQNN-TE-QPINN arrives only at epoch 77 on average. Total improvement from epoch 1 to epoch 100 follows the same order, $2.7\%$ for Golomb, $31.3\%$ for QPINN, $52.1\%$ for TowerCheb and $57.8\%$ for AdaFreq-QPINN, against $88.1\%$ for LQNN-TE-QPINN, which descends from $0.7672$ to $0.0916$. Between epoch 20 and epoch 100 the identity baseline gains only $0.0118$ and TowerCheb $0.0343$, while LQNN-TE-QPINN gains $0.1588$, from $0.2504$ to $0.0916$. 
A plausible explanation is that the predefined encoding frequencies
interact less favorably with the target solution and subsequent
variational optimization at this configuration, leading to earlier
plateaus. The training-budget results further show that these plateaus
are not removed simply by increasing the number of optimization
epochs over the tested range. AdaFreq-QPINN provides an informative
intermediate case: it retains the analytical arccos-Chebyshev form and
promotes only the encoding-frequency scales $\lambda_k$, shared by each
wire pair in this one-dimensional setting, to trainable parameters,
finishing as the best analytical embedding at $0.3262$ with only four
additional trainable parameters. This result indicates that adapting
the encoding-frequency scales can improve performance over the evaluated
fixed analytical embeddings at this configuration. AdaFreq-QPINN
nevertheless reaches its plateau at epoch 16 and finishes $3.6\times$
behind LQNN-TE-QPINN at this configuration. The two forms of adaptivity
therefore exhibit different approximation and optimization behavior
here, with LQNN-TE-QPINN continuing to improve over a larger fraction
of the training budget. We do not interpret this result as establishing
a general hierarchy between the two approaches, since the
ordering depends on the training budget: the two-dimensional comparison
in Figure~\ref{fig:2d-embedding} places their solution errors within one
standard deviation of each other at 100 epochs, whereas the 200-epoch
repetition in Table~\ref{tab:2d200} separates them in favor of
LQNN-TE-QPINN.

\begin{figure*}[t]
\centering
\includegraphics[width=\textwidth]{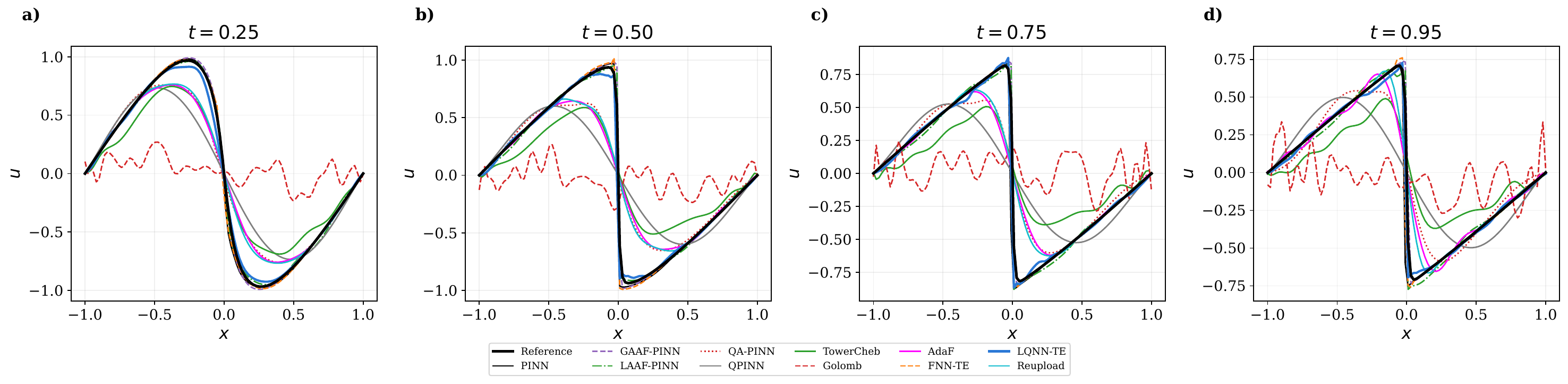}
\caption{Reconstructed solution profiles $u(x)$ of the eleven models trained at the anchor configuration, a main VQC of 8 qubits and 15 layers with a 4-qubit, 30-layer trainable embedding, reloaded from their saved parameters and evaluated on the $100 \times 100$ grid over seeds 42, 123 and 2024. The RK45 reference is the thick black line and LQNN-TE-QPINN is the blue solid curve. (a) $t = 0.25$; (b) $t = 0.50$; (c) $t = 0.75$; (d) $t = 0.95$.}
\label{fig:inference}
\end{figure*}

One reading of Figure~\ref{fig:qpinn_lqnn}(b) requires pre-emption, since four of the five bands are hairlines and only LQNN-TE-QPINN carries a wide envelope, inviting the conclusion that the analytical embeddings are the reliable ones. Averaged over epochs 20 to 100, the seed spread as a fraction of the mean is $0.17\%$ for QPINN, $0.77\%$ for Golomb, $0.95\%$ for AdaFreq-QPINN, $6.7\%$ for TowerCheb and $21.7\%$ for LQNN-TE-QPINN. 
Those tight bands should not by themselves be interpreted as
evidence of greater reliability, since the corresponding models
reach early plateaus under this configuration and therefore exhibit
little subsequent variation across seeds.
The band therefore carries information only for LQNN-TE-QPINN, and we read the panel as a comparison of mean curves with a single informative envelope. The same model behaves oppositely against the classical PINN baselines of Figure~\ref{fig:pinn_lqnn}, where its final spread of $\pm 0.0021$ is the tightest of that group, so seed sensitivity is a property of a model at a given configuration and never a fixed trait. Two limitations remain. First, the frequency-tower members are distinguished only by the integer sequence $\{\lambda_k\}$, which is truncated to its first $P = \lceil n/(d+1) \rceil$ entries, so the qubit budget that classical simulation permits also bounds how many of these embeddings stay distinguishable. In one dimension, where $P = \lceil n/2 \rceil$, at 2 qubits all five collapse onto $\{1\}$ and are literally the same model, at 4 qubits the five names reduce to two distinct models, since TowerCheb, Binary and Hamming all realize $\{1,2\}$ while Golomb and Turnpike both realize $\{1,3\}$, and at 6 qubits Binary and Hamming still coincide at $\{1,2,4\}$. At the 8-qubit anchor of this figure the five sequences are genuinely distinct, so we attribute the omission of Turnpike, Binary and Hamming to the compute budget of this reduced run and not to any degeneracy. We mitigate that omission by evaluating the missing members where they matter, since Turnpike and Hamming enter the two-dimensional comparison of Figure~\ref{fig:2d-embedding}, where the coordinate triples give $P = \lceil 5/3 \rceil = 2$, so that TowerCheb, Binary and Hamming all realize $\{1,2\}$ while Golomb and Turnpike both realize $\{1,3\}$, and the two towers plotted there are precisely the two distinct fixed encodings that this width admits. Second, the evidence rests on one PDE at one anchor configuration, which we mitigate through the sizing results above, which fix 8 qubits and 15 layers as the anchor for this comparison, and through the embedding taxonomy of Section~IV, which characterizes the omitted members analytically.

\subsubsection{Comparison against Alternative Hybrid Architectures}
Figure~\ref{fig:others} places the LQNN-TE-QPINN against three alternative solver architectures at one anchor, a main VQC of 8 qubits and 15 layers with the trainable embedding held at 4 qubits and 30 layers, so that the architecture is the only variable, and we train every model for 100 L-BFGS epochs over three seeds (42, 123 and 2024) and draw the mean curve with a band of one standard deviation. FNN-TE-QPINN keeps a trainable embedding but generates the encoding angles with a classical feed-forward network in place of the auxiliary quantum circuit. QA-PINN keeps a classical PINN and substitutes a single hidden layer with a quantum circuit, so the quantum component sits inside a classical network instead of acting as the feature map. Reupload-QPINN removes the intermediate readout entirely, passing no classical angles forward and re-uploading the coordinates inside the solver circuit, so every parameter is optimized through one quantum state. At epoch 100 the proposed model attains the lowest relative $L_2$ error of the four, $0.0916 \pm 0.0021$ with 720 trainable parameters, which is $1.6\times$ more accurate than FNN-TE-QPINN ($0.1456 \pm 0.0493$) at $1.2\times$ the parameters, $3.3\times$ more accurate than Reupload-QPINN ($0.2979 \pm 0.0003$) at $2\times$ the parameters, and $3.4\times$ more accurate than QA-PINN ($0.3076 \pm 0.0062$) at $1.2\times$ the parameters. Its mean curve passes QA-PINN at epoch 4 and Reupload-QPINN at epoch 5 and stays below both thereafter, yet it settles permanently below FNN-TE-QPINN only at epoch 71, and that late crossing requires an explanation.

The classical trainable embedding in fact leads through the middle of the run. The two are level at epoch 40, $0.1981$ against $0.2023$, and FNN-TE-QPINN leads at epoch 60, $0.1511$ against $0.1662$, after which the proposed model descends to $0.1195$ at epoch 80 and $0.0916$ at epoch 100 while its classical counterpart stalls at $0.1423$ and regresses to $0.1456$. The training loss ranks the pair in the opposite order, $0.0097$ against $0.0212$, so selecting by the objective would return the model carrying $59\%$ more solution error for $18\%$ fewer parameters. This reversal may be attributed to over-training in the classical embedding. Measured from its own best epoch to epoch 100, FNN-TE-QPINN degrades by $1.5\%$, $4.3\%$ and $22.4\%$ on seeds 42, 123 and 2024, with best epochs at 81, 94 and 45, whereas the proposed model degrades by $0.0\%$, $0.5\%$ and $3.8\%$ with best epochs at 100, 95 and 61. On seed 2024 it bottoms out at epoch 45 and then surrenders nearly a quarter of its accuracy while its training loss keeps falling. We qualify the reliability claim immediately: averaged over epochs 20 to 100 the seed spread as a fraction of the mean is $22.2\%$ for FNN-TE-QPINN against $21.7\%$ for the proposed model, so neither is steadier during training and a mid-training snapshot would misrepresent both. What separates them is the converged state, where the standard deviation is $\pm 0.0493$ against $\pm 0.0021$, a factor of 23.

\begin{figure*}[t]
\centering
\includegraphics[width=0.9\textwidth]{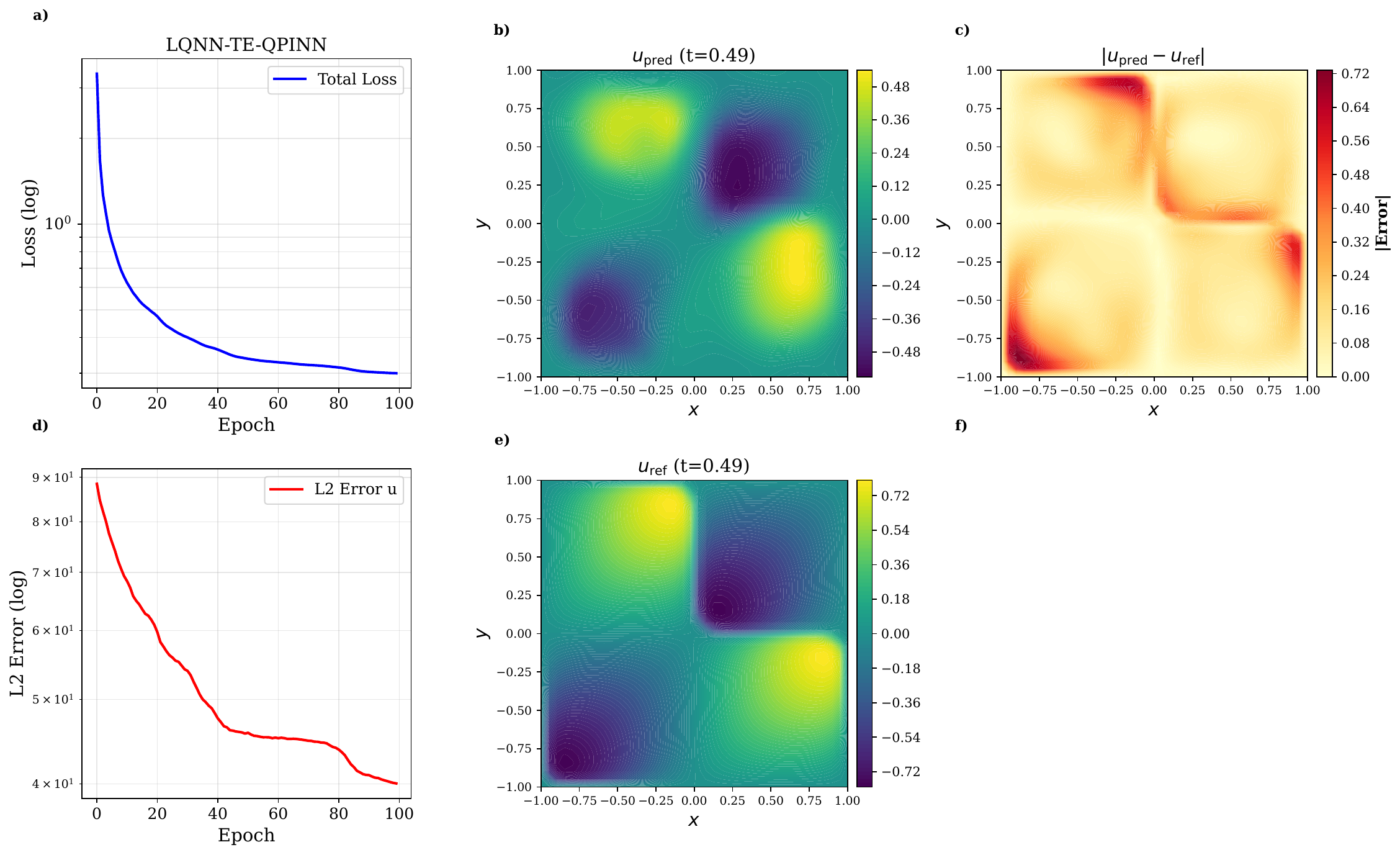}
\caption{Training performance of LQNN-TE-QPINN on the two-dimensional Burgers equation after 100 epochs, evaluated with a main VQC of 5 qubits and 15 layers and a trainable embedding of 4 qubits and 30 layers, the embedding setting the one-dimensional sizing identified as best. Collocation points were subsampled to a grid resolution of $40 \times 40 \times 40$ to prevent out-of-memory errors. (a) and (d) Total PDE loss and $L_2$ error of the solver, averaged over seeds 42, 1004 and 2026. (b) and (e) Heatmap comparison of the predicted solution $u$ versus the reference solution at $t = 0.49$. (c) Absolute error between the predicted and reference solutions.}
\label{fig:2d-training}
\end{figure*}

Reupload-QPINN carries a result that invites a misreading, since it is the most reproducible model in the study at $\pm 0.0003$, seven times tighter than the $\pm 0.0021$ of the proposed model, while sitting $3.3\times$ higher in relative $L_2$ error on half as many parameters. It reaches within $5\%$ of its final error at epoch 10 on average, against epoch 77 for the proposed model. A model that stops improving early and then holds its value is trivially consistent, so consistency across seeds is not by itself evidence of a good model. QA-PINN repeats the pattern in milder form at $\pm 0.0062$ and epoch 66, moving only from $0.3703$ at epoch 20 to $0.3076$ at epoch 100. A plausible explanation is that a single quantum hidden layer inside an otherwise classical network leaves the representation dominated by the classical layers, so the architecture inherits the convergence profile of the classical baseline instead of that of a quantum feature map. These pairs indicate that the property worth claiming for the proposed model is a conjunction, the lowest error together with a spread that collapses at convergence, since either half alone is satisfied here by a model three times worse. The comparison nevertheless has limitations, as it holds at one anchor and one PDE. We tuned each alternative no further than the proposed model, and FNN-TE-QPINN reached the lower training loss, so we do not claim that it is the weaker approximator; the observation is that at this configuration its lower physics-informed loss does not translate into a lower solution error. A longer budget would test whether the over-training persists and whether Reupload-QPINN moves at all after its plateau.

\begin{table}[!t]
\caption{The ten models of this section at the common one-dimensional anchor, a main VQC of 8 qubits by 15 layers with a 4-qubit, 30-layer trainable embedding, trained for 100 L-BFGS epochs on the $100\times100$ grid, mean $\pm$ standard deviation over seeds 42, 123 and 2024. Rows are ordered by decreasing solution error. The last column gives the epoch at which each model first comes within $5\%$ of its own final relative $L_2$ error, averaged over the three seeds. The values consolidate Figs.~\ref{fig:pinn_lqnn}--\ref{fig:others}, which report the same run.}
\label{tab:1danchor}
\centering
\footnotesize
\setlength{\tabcolsep}{3.5pt}
\renewcommand{\arraystretch}{1.2}
\begin{tabular}{|l|r|r@{\,$\pm$\,}l|r@{\,$\pm$\,}l|r|}
\hline
\textbf{Model} & \textbf{Par.} & \multicolumn{2}{c|}{\textbf{Loss}} & \multicolumn{2}{c|}{\textbf{Rel.\ $L_2$}} & \textbf{Ep.} \\
\hline
Golomb         & 360  & 19.819 & 1.037 & 0.9452 & 0.0074 & 1 \\
QPINN          & 360  & 0.432 & 0.002 & 0.4501 & 0.0013 & 7 \\
TowerCheb      & 360  & 0.343 & 0.051 & 0.3684 & 0.0293 & 27 \\
AdaFreq-QPINN  & 364  & 0.260 & 0.002 & 0.3262 & 0.0020 & 16 \\
QA-PINN        & 600  & 0.279 & 0.008 & 0.3076 & 0.0062 & 66 \\
Reupload-QPINN & 360  & 0.245 & 0.000 & 0.2979 & 0.0003 & 10 \\
PINN           & 7851 & 0.102 & 0.124 & 0.1751 & 0.0981 & 71 \\
FNN-TE-QPINN   & 588  & \textbf{0.010} & \textbf{0.009} & 0.1456 & 0.0493 & 58 \\
LAAF-PINN      & 8051 & 0.053 & 0.041 & 0.1233 & 0.0739 & 33 \\
LQNN-TE-QPINN  & 720  & 0.021 & 0.013 & \textbf{0.0916} & \textbf{0.0021} & 77 \\
\hline
\end{tabular}
\end{table}

Table~\ref{tab:1danchor} collects the ten models at this common anchor, so that the four comparisons above are read against one another rather than figure by figure. Two points stand out. The physics-informed objective and the solution error do not rank the models alike: FNN-TE-QPINN attains the lowest training loss of the ten, $0.010$, yet places third by solution error, while the proposed model carries twice that loss and the lowest error, and Golomb reports a loss three orders of magnitude above every other entry on an error of $0.9452$. Selecting by the objective would therefore return a different model at this anchor, which is the reason we let the relative $L_2$ error carry every ranking in this work. The parameter column adds the second point: the two lowest errors belong to LQNN-TE-QPINN at 720 parameters and LAAF-PINN at 8,051, so the quantum model reaches the better value with roughly eleven times fewer. The final column should be read with care, since settling late does not by itself indicate a better model: QA-PINN is slow to settle, at epoch 66, and still finishes mid-table. What it does show is the converse, that every model finishing below $0.18$ keeps improving past epoch 30, whereas the four fixed or re-uploading encodings reach their final value within the first sixteen epochs and spend the rest of the budget without moving.

\subsubsection{Inferences}
We reload the saved parameters of the eleven trained models and evaluate them on the $100 \times 100$ grid of the one-dimensional Burgers problem, 10000 points of which 100 are initial, 198 boundary, and 9702 interior, with the main VQC at 8 qubits by 15 layers, the trainable embedding at 4 qubits by 30 layers, and three seeds (42, 123, 2024). The reload serves first as a round-trip check on the checkpoint path, since every relative $L_2$ value recovered from the saved weights matches the corresponding training run to four significant figures, so the rebuilt circuits agree with the trained models and nothing in the profiles below is an artifact of loading. It also yields the maximum pointwise deviation $L_\infty$, a worst-case quantity the training curves never report. LQNN-TE-QPINN attains the lowest $L_\infty$ at $0.633 \pm 0.015$, alongside the lowest relative $L_2$ at $0.0916 \pm 0.0021$ and the tightest seed spread on both. Among the models that train reliably, LQNN-TE-QPINN therefore holds the lowest worst-case error and the most stable one. Two models exceed $1.0$ in $L_\infty$, FNN-TE-QPINN at $1.136$ and Golomb at $1.147$, and since such a value exceeds the amplitude of the reference solution itself, both overshoot somewhere in the domain. FNN-TE-QPINN reaches this despite placing third in relative $L_2$ at $0.1456$, so the two norms disagree about it, repeating between error metrics the objective-versus-solution divergence that the preceding subsections establish.

Figure~\ref{fig:inference} illustrates the reconstructed profiles $u(x)$ at (a) $t = 0.25$, (b) $t = 0.50$, (c) $t = 0.75$, and (d) $t = 0.95$, with the RK45 reference drawn as a thick black line and LQNN-TE-QPINN as the blue solid curve. At $t = 0.25$ the solution remains smooth and most models track it, although the identity-encoding QPINN already flattens the negative lobe visibly. As the convective term steepens the front toward $x = 0$ at $t = 0.75$ and $t = 0.95$, the separation becomes clear: LQNN-TE-QPINN follows the steep transition on both sides and stays close to the reference, the identity-encoding QPINN smooths the interface into a rounded profile that underestimates the amplitude near the shock, and Golomb oscillates about zero at every instant across the whole domain, consistent with its relative $L_2$ of $0.9452$, which it already reached at the first training epoch. Two limitations bound this evaluation. The evaluation grid is the training grid, both linearly spaced over the same domain, so this subsection establishes reconstruction quality and worst-case error and does not establish generalization to unseen points. Binary and Hamming are also absent, because this anchor run did not train them. A held-out grid at a different resolution would convert the reconstruction check into a generalization test at the cost of one evaluation pass, since no retraining is involved.

\subsection{Two-Dimensional Training Evaluation}

\begin{figure*}[t]
\centering
\includegraphics[width=0.9\textwidth]{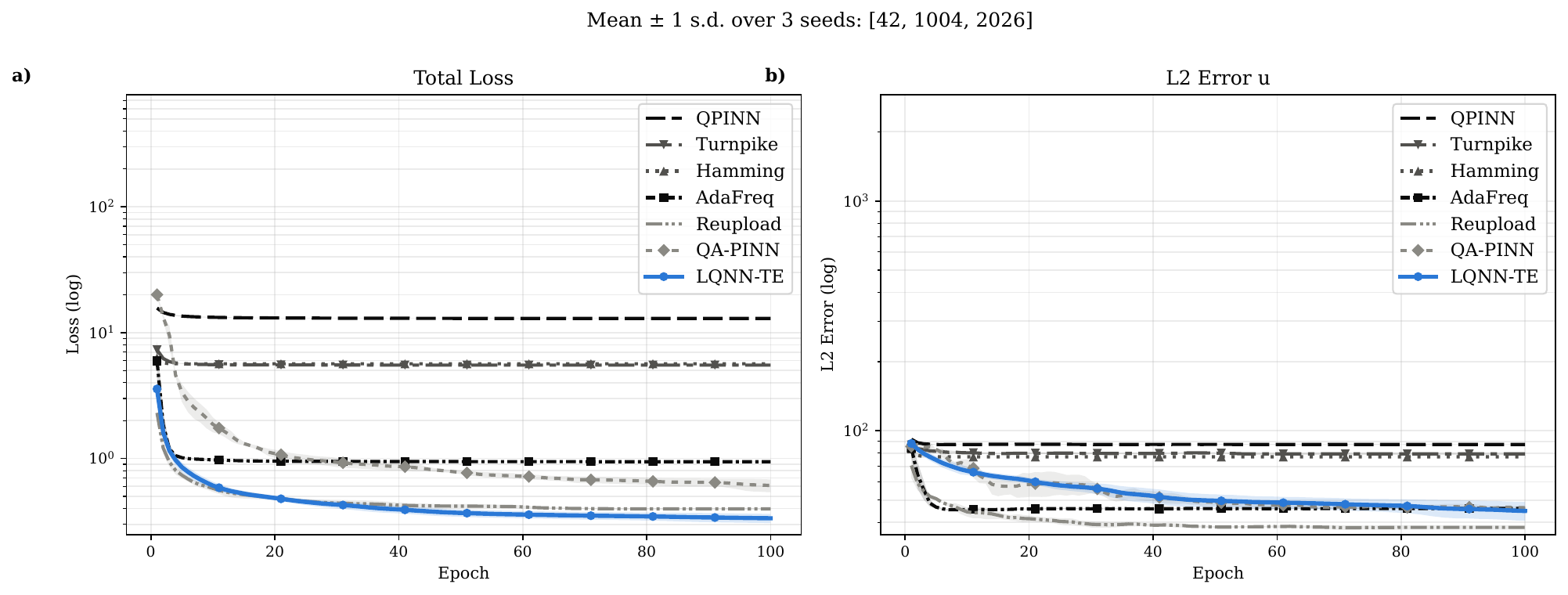}
\caption{Performance evaluation of the LQNN-TE-QPINN solver on the two-dimensional Burgers equation against six alternative embedding methods and architectures, all at the same configuration of a main VQC of 5 qubits and 15 layers with a 4-qubit, 30-layer trainable embedding, trained for 100 epochs over seeds 42, 1004 and 2026. (a) Solver loss function progression during training. (b) $L_2$ error of the solver evaluated against the RK45 reference solution.}
\label{fig:2d-embedding}
\end{figure*}

The error metric differs between the two benchmarks and the two sets of numbers must not be compared in magnitude. The one-dimensional experiments above report the relative $L_2$ error $\lVert \tilde{u}-u_{\mathrm{ref}}\rVert_2 / \lVert u_{\mathrm{ref}}\rVert_2$, which is dimensionless, whereas the two-dimensional implementation reports the unnormalized discrete $L_2$ norm $\lVert \tilde{u}-u_{\mathrm{ref}}\rVert_2$ evaluated on the collocation grid. The two-dimensional values below are therefore meaningful only relative to one another, and we refer to them as the solution error rather than as a relative $L_2$ error.

We solve the two-dimensional problem over $(x,y)\in[-1,1]^2$ and
$t\in[0,0.95]$ at the same viscosity $\nu=0.01/\pi$, and we compute its
reference solution with the same RK45 method-of-lines scheme, using
first-order upwind differences for both advection terms on the
collocation grid.

To assess the scalability of the proposed approach to higher dimensions, we extend the evaluation to the two-dimensional Burgers equation, fixing the main variational quantum circuit at 5 qubits by 15 layers and holding the trainable embedding at the 4 qubit by 30 layer setting that the one-dimensional sizing identified as best. 
At the five-qubit two-dimensional configuration, the number of
frequency-scaling factors is
$
P=\left\lceil \frac{n}{d+1}\right\rceil
=\left\lceil \frac{5}{3}\right\rceil
=2.
$
Consequently, TowerCheb, Binary, and Hamming share the same
scaling sequence $\{1,2\}$, while Golomb and Turnpike share
$\{1,3\}$. We therefore report Hamming and Turnpike as
representatives of the two distinct fixed analytical embedding
configurations at this circuit width.

Five qubits is what the two-dimensional collocation set permits under classical simulation, and for the same reason we subsample the domain to a $40 \times 40 \times 40$ grid of 64000 points, 1600 initial, 6084 boundary, and 56316 interior. We train every model for 100 epochs over three seeds (42, 1004, and 2026). Figure~\ref{fig:2d-training} presents the proposed LQNN-TE-QPINN alone, where panels (a) and (d) indicate a steady and smooth decline of the total PDE loss from $3.57$ at epoch 1 to $0.3349$ at epoch 100, crossing below the order of $10^0$ within the first ten epochs at $0.6095$, with the $L_2$ error following the same trend and without the early saturation the fixed encodings enter. The heat maps in panels (b) and (e) demonstrate that the predicted field at $t = 0.49$ reproduces the structure of the reference solution, and the absolute error in panel (c) concentrates where the solution varies most sharply, which is the behavior we expect of a residual-driven solver carrying 585 trainable parameters at this budget.

Figure~\ref{fig:2d-embedding} compares all seven models. At epoch 100 the LQNN-TE-QPINN attains an $L_2$ error of $44.82 \pm 4.22$ and ranks second, since Reupload-QPINN reaches $38.03 \pm 0.83$ with 225 parameters against our 585 and leads outright. AdaFreq-QPINN follows at $45.93 \pm 0.15$, QA-PINN at $46.26 \pm 1.12$, Hamming at $77.05$, Turnpike at $79.39$, and the identity-encoding QPINN baseline at $87.19$. The proposed model nevertheless records the lowest final objective of the seven, $0.3349$ against $0.3980$ for Reupload-QPINN, so this setting reproduces the objective-versus-solution divergence the one-dimensional sections establish, and we let the $L_2$ error carry the ranking as before. The mean curve of the proposed model falls permanently below QPINN from epoch 1, Turnpike from epoch 2, Hamming from epoch 4, QA-PINN from epoch 84, and AdaFreq-QPINN from epoch 88, while Reupload-QPINN is never overtaken within the budget.
Within this experiment, Reupload-QPINN, LQNN-TE-QPINN,
AdaFreq-QPINN, and QA-PINN form the lower-error group,
with solution errors ranging from $38.03$ to $46.26$, whereas
Hamming, Turnpike, and the direct-encoding QPINN baseline
produce substantially larger errors.
We do not compare the size of this separation with the one-dimensional one, since the two subsections report different error metrics.
Those three models reach their final level within 2 to 3 epochs,
with Hamming remaining nearly unchanged thereafter. This behavior
indicates early plateaus for the evaluated fixed embeddings at this
configuration, consistent with the saturation behavior observed in
the one-dimensional analytical comparison.

The training budget appears to be the binding factor: the epoch at which each model first comes within $5\%$ of its own final $L_2$ error, on average, is 2 for QPINN, 2 for Hamming, 3 for Turnpike, 5 for AdaFreq-QPINN, 32 for Reupload-QPINN, 57 for QA-PINN, and 73 for the proposed model, the slowest of the seven to settle. Its mean error is still falling steeply at the close of training, from $52.08$ at epoch 40 to $48.80$, $47.24$, and $44.82$ at epochs 60, 80, and 100, and two of its three overtakes fall in the final fifth of the run, whereas Reupload-QPINN moves only from $38.32$ at epoch 60 to $38.03$ at epoch 100. The ordering therefore appears to be taken where one model has converged and the other has not, so 100 epochs may be too short to separate them. The proposed model also carries the largest spread of the seven at $\pm 4.22$, driven by seed 1004 at $50.31$ against $40.05$ and $44.10$ on the other two, where Reupload-QPINN holds $\pm 0.83$ and AdaFreq-QPINN $\pm 0.15$. This reverses the one-dimensional picture, in which the proposed model held the tightest converged spread of its group, and supports our earlier principle that seed sensitivity is a property of a model at a given configuration and never a fixed trait. Two limitations therefore stand: the proposed model has not clearly converged within the tested budget, so its reported error should not be read as a converged value, and the qubit count and grid resolution remain dictated by classical simulation cost. The embedding is at least no longer a confound, since we hold it at the setting the one-dimensional sizing identified as best. We therefore repeated the experiment at the same configuration for 200 epochs, and report it as an extension of the analysis above rather than as a replacement for it.

\begin{table}[!t]
\caption{Two-dimensional Burgers at 200 epochs, same configuration as Figs.~\ref{fig:2d-training} and \ref{fig:2d-embedding}: a main VQC of 5 qubits by 15 layers with a 4-qubit, 30-layer trainable embedding on a $40\times40\times40$ grid, mean $\pm$ standard deviation over seeds 42, 1004 and 2026. The solution error is the unnormalized discrete $L_2$ norm defined at the start of this subsection. Model names are abbreviated: AdaFreq, Reupload and LQNN-TE denote AdaFreq-QPINN, Reupload-QPINN and LQNN-TE-QPINN. The last column gives the mean solution error at epoch 100 within this same run, so that the trajectory is read without mixing runs.}
\label{tab:2d200}
\centering
\footnotesize
\setlength{\tabcolsep}{3.5pt}
\renewcommand{\arraystretch}{1.2}
\begin{tabular}{|l|r|r@{\,$\pm$\,}l|r@{\,$\pm$\,}l|r|}
\hline
\textbf{Model} & \textbf{Par.} & \multicolumn{2}{c|}{\textbf{Loss}} & \multicolumn{2}{c|}{\textbf{Error}} & \textbf{Ep.100} \\
\hline
QPINN     & 225 & 12.949 & 0.013 & 87.19 & 0.14 & 87.19 \\
Turnpike  & 225 & 5.517 & 0.005 & 79.39 & 0.89 & 79.39 \\
Hamming   & 225 & 5.657 & 0.000 & 77.05 & 0.00 & 77.05 \\
AdaFreq   & 227 & 0.941 & 0.004 & 45.94 & 0.14 & 45.93 \\
QA-PINN   & 605 & 0.548 & 0.072 & 44.93 & 1.84 & 46.26 \\
Reupload  & 225 & 0.407 & 0.009 & \textbf{39.11} & \textbf{1.42} & 38.89 \\
LQNN-TE   & 585 & \textbf{0.298} & \textbf{0.009} & 40.58 & 2.18 & 44.82 \\
\hline
\end{tabular}
\end{table}

Table~\ref{tab:2d200} confirms three findings of the 100-epoch analysis. The ranking by solution error is unchanged, with Reupload-QPINN lowest at $39.11$ and the proposed model second at $40.58$. The proposed model again attains the lowest physics-informed objective of the seven, $0.298$ against $0.407$, so the divergence between the training objective and the solution error persists at the longer budget and is not an artifact of stopping early. The three fixed encodings again do not move, returning the same values at epoch 200 as at epoch 100 to every digit recorded. The AdaFreq-QPINN parameter count also confirms the grouping rule of Section~IV: at five qubits in three input dimensions the encoding uses $P = \lceil 5/3 \rceil = 2$ trainable frequency scales, which is the two-parameter difference between its 227 and the 225 of the fixed towers.

The doubled budget sharpens the picture in one direction and revises it in two others. It sharpens the trajectory, since the proposed model improves by $9.5\%$ between epoch 100 and epoch 200, from $44.82$ to $40.58$, while no other model moves by more than $3\%$, so the gap to Reupload-QPINN closes from $5.93$ to $1.47$ and the proposed model wins outright on seed 42, $38.08$ against $41.11$. It revises, first, the reading of seed sensitivity: the spread of the proposed model halves from $\pm 4.22$ to $\pm 2.18$ while that of Reupload-QPINN widens from $\pm 0.83$ to $\pm 1.42$, so the proposed model no longer carries the largest spread of the seven. It revises, second, our description of Reupload-QPINN as a model that stops improving early and then holds its value: on seed 42 it reaches $38.57$ at epoch 61 and then degrades by $6.6\%$ to $41.11$, over-training at this budget in the manner we attributed to FNN-TE-QPINN in one dimension.

The longer budget nevertheless does not settle the ordering, contrary to what the 100-epoch discussion anticipated, and for two reasons that we state directly. The first is that doubling the budget does not supply the same additional training to every model. Measured by the last epoch at which L-BFGS still updates the parameters, Hamming stops at epoch 19, Turnpike at 56, Reupload-QPINN at 86 and QPINN at 96 on average, whereas the proposed model is still updating at epoch 200 and records its best value at epochs 198, 200 and 200 on the three seeds. The comparison at epoch 200 therefore places a model that has stopped against one that has not, and we do not extrapolate the remaining gap.

The second reason is a reproducibility limit that bears directly on the ranking. Comparing the two runs cell by cell at epoch 100, six of the seven models agree to every recorded digit on all three seeds, so the last column of Table~\ref{tab:2d200} may be read against Fig.~\ref{fig:2d-embedding} directly. Reupload-QPINN is the exception, reading $38.89$ in the last column of Table~\ref{tab:2d200} against the $38.03$ that the figures of this subsection report at the same epoch, a difference of $1.40$ on seed 42 and $1.20$ on seed 1004 with the two trajectories separating as early as epoch 3 at a loss difference of order $10^{-6}$. The assignment of cells to devices was identical in the two runs, so we attribute this to non-deterministic floating-point reduction, amplified by the L-BFGS line search, in what is the deepest single tape among the models compared here. The consequence is quantitative: the $1.47$ separating Reupload-QPINN from the proposed model at epoch 200 is of the same order as the $0.87$ mean run-to-run drift of Reupload-QPINN itself and as its $\pm 1.42$ seed spread. We therefore do not claim that Reupload-QPINN is the better model at this configuration. We claim only that the two are not separated by the evidence available at this budget and grid resolution, and that a deterministic re-run of the two-dimensional comparison is the experiment this calls for.

\subsection{Noise Evaluation}

\begin{figure*}[t]
\centering
\includegraphics[width=\textwidth]{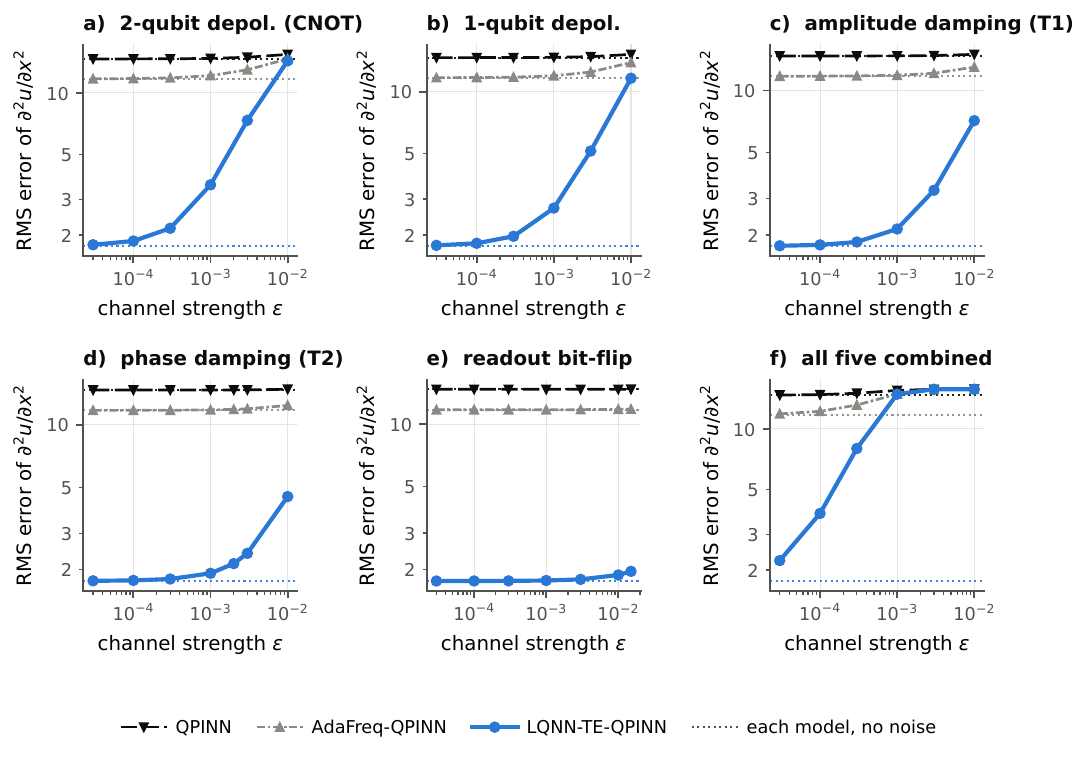}
\caption{Impact of five quantum noise channels, isolated one at a time, on QPINN, AdaFreq-QPINN and LQNN-TE-QPINN at the anchor configuration. Each panel plots the total RMS error of $\partial^{2}u/\partial x^{2}$ against the channel strength $\varepsilon$, with the zero-noise error of each model drawn as a dotted line in its own color, so that both the ranking and the amount each channel adds remain visible. (a) 2-qubit depolarizing (CNOT), (b) 1-qubit depolarizing, (c) amplitude damping, (d) phase damping, (e) readout bit-flip, (f) all five applied together. Panels (a) and (b) differ mainly in how many times a channel is inserted, 100 against 66 at this width, so their ordering reflects circuit topology; (c) and (d) share both position and insertion count and are the one strictly matched pair.}
\label{fig:noise}
\end{figure*}

\begin{figure*}[t]
\centering
\includegraphics[width=0.8\textwidth]{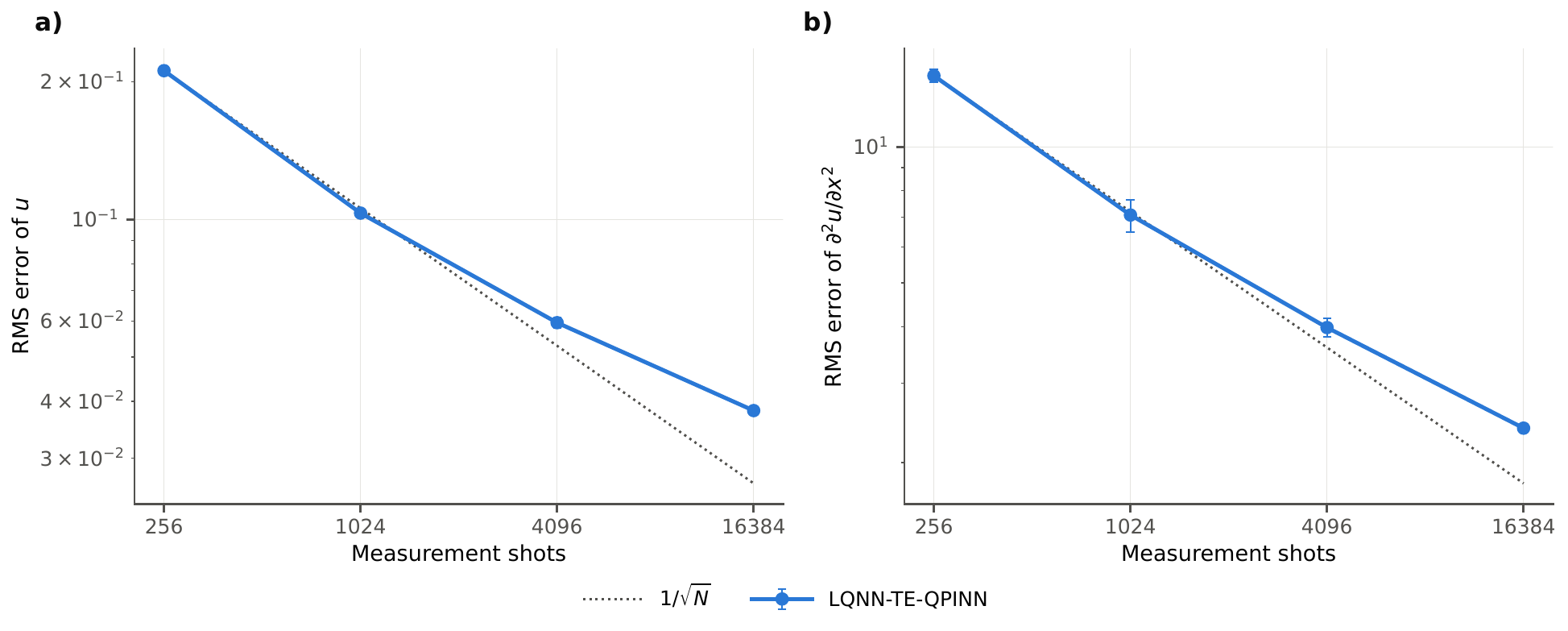}
\caption{Evaluation of finite-shot noise effects on the LQNN-TE-QPINN model. (a) RMS error of $u$. (b) RMS error of $\partial^{2}u/\partial x^{2}$. The shot count ranges from 256 to 16,384, and the dotted line indicates the $1/\sqrt{N}$ scaling of shot noise.}
\label{fig:finite-shot}
\end{figure*}

Figure~\ref{fig:noise} evaluates the three solvers under five hardware-like noise channels, isolated one at a time at the anchor of a main VQC of 8 qubits by 15 layers with a trainable embedding of 4 qubits by 30 layers, plotting the total root-mean-square (RMS) error of $\partial^{2}u/\partial x^{2}$ against strength $\varepsilon$ on logarithmic axes. At zero noise the LQNN-TE-QPINN starts at $1.764$ against $11.72$ for AdaFreq-QPINN and $14.70$ for the identity baseline, margins of $6.6\times$ and $8.3\times$ obtained without widening the solver circuit. Two distinct quantities must be kept apart throughout this subsection. The first is the absolute error a model attains under a given channel at a given strength, which decides the ranking. The second is the degradation of a model relative to its own noiseless baseline, which is what noise robustness properly denotes. The two need not agree, and in this study they do not. Across the whole swept range the proposed model holds the lowest total error in every individual channel, and at the strongest strength swept it records $14.386$, $11.657$, $7.147$ and $4.505$ under 2-qubit depolarizing (CNOT), 1-qubit depolarizing, amplitude damping and phase damping at $\varepsilon = 10^{-2}$, against $14.704$, $13.908$, $12.985$ and $12.385$ for AdaFreq-QPINN and $15.471$, $15.241$, $14.957$ and $14.821$ for the baseline, together with $1.963$ against $11.808$ and $14.708$ under readout bit-flip at $1.5\times10^{-2}$. This severity ordering, with readout mildest at $1.886$ for $\varepsilon = 10^{-2}$, carries two qualifications. The gap between the first two channels is largely gate counting, since the 1-qubit channel is inserted 66 times and the 2-qubit depolarizing (CNOT) channel 100 times, a ratio of $1.515$, against measured added-error ratios of $1.523$, $1.492$ and $1.452$, so insertion counting reproduces it within $5\%$, and this part of the ordering describes circuit topology and not channel physics. Amplitude damping and phase damping share both position and insertion count, forming the one strictly matched pair, and phase damping proves the milder ($4.505$ against $7.147$).

The combined channel of Figure~\ref{fig:noise}(f), with all five sources active, marks where this advantage ends. It records $2.228$ at $\varepsilon = 3\times10^{-5}$ and $3.813$ at $10^{-4}$ against $11.873$ and $12.224$ for AdaFreq-QPINN and $14.721$ and $14.792$ for the baseline, and retains a clear margin at $3\times10^{-4}$ with $7.999$ against $13.140$ and $15.016$. At $10^{-3}$ the three models sit at $14.863$, $14.881$ and $15.521$, so the advantage is gone, and at $10^{-2}$ all three reach $15.720$, a regime where the choice of model stops carrying information. The underlying weakness is real. In a separate six-qubit study at depolarizing strength $p = 0.05$, the added error grows with derivative order and fastest for the proposed model: QPINN gains $0.284$, $0.520$ and $0.643$ in $u$, $\partial u/\partial x$ and $\partial^{2}u/\partial x^{2}$, an amplification of $2.3\times$, AdaFreq-QPINN gains $0.352$, $0.933$ and $2.980$, an amplification of $8.5\times$, and LQNN-TE-QPINN gains $0.389$, $1.277$ and $5.756$, an amplification of $14.8\times$. This behavior may stem from the architecture itself. We verified in the implementation that the auxiliary embedding circuit and the solver circuit are both executed on the noisy simulator under the same channel specification, so for LQNN-TE-QPINN the noise is applied to two circuits rather than one and the effective depth of a single model evaluation is larger. The channels are also inserted once per layer rather than once per circuit, so their contribution accumulates with depth in both circuits.

The two quantities separated above therefore point in opposite directions, and the anchor experiment shows this without recourse to the six-qubit study. Measured against its own noiseless value, LQNN-TE-QPINN degrades by a factor of $8.2$ under 2-qubit depolarizing (CNOT) at $\varepsilon = 10^{-2}$ and $8.4$ under the combined channel at $10^{-3}$, whereas AdaFreq-QPINN degrades by $1.25$ and $1.27$ and the identity baseline by $1.05$ and $1.06$. On this measure the proposed model is the least robust of the three. We read the two baseline figures with caution, however, since a noiseless error of $14.70$ already sits close to the level of $15.72$ at which all three models saturate, so that model has little room in which to degrade and its small ratio measures the absence of headroom as much as any resistance to the channels. What the comparison does establish is that the lowest absolute error and the smallest relative degradation are not held by the same model. That LQNN-TE-QPINN nonetheless holds the lowest absolute error at every strength tested, until the combined channel saturates all three, follows from its starting far below the others rather than from any insensitivity to noise. We therefore claim the lowest absolute derivative error under the simulated noise settings considered here, and we do not claim intrinsic noise robustness.

Figure~\ref{fig:finite-shot} isolates the finite measurement budget for the LQNN-TE-QPINN at the same anchor over three repeats. The RMS error of $u$ falls from $0.2117 \pm 0.0015$ at 256 shots to $0.0382 \pm 0.0005$ at 16384 against an exact-expectation floor of $0.0287$, and the second derivative from $14.3834 \pm 0.4855$ to $2.3827 \pm 0.0264$ against a floor of $1.7640$. Against each metric's own floor, 256 shots multiply the error by $7.37$, $7.93$ and $8.15$ for the three derivative orders and 16384 shots by $1.33$, $1.35$ and $1.35$, so sampling degrades every order by nearly the same factor, unlike the channel noise above, and the excess decays consistently with the inverse square root scaling of shot noise. Quoting 16384 shots brings the second derivative within $35\%$ of its noiseless value at unchanged circuit width. Model comparison under shots exists only in the separate six-qubit study, where 256 shots raise the RMS error of $u$ by $18\%$ for QPINN, $25\%$ for AdaFreq-QPINN and $66\%$ for LQNN-TE-QPINN ($0.1718$ to $0.2855$), and the ranking holds at every budget. Two limitations qualify these figures: a depolarizing-only noise model underestimates severely, since at a matched 1-qubit error rate of $10^{-3}$ the proposed model gains $0.0132$ in the RMS error of $u$ under depolarizing alone against $0.2442$ under the full hardware-like specification, a factor of $18.6$, so we quote hardware-like numbers throughout. Furthermore, we implement the 2-qubit depolarizing (CNOT) channel as single-qubit depolarizing channels of equal strength on both operands of every CNOT gate, so it captures gate-associated error on the two qubits without modeling correlated two-qubit error. We mitigate both, since isolating the channels keeps their contributions separable, the matched damping pair gives one comparison free of the gate-counting confound, and the shot study fixes the budget at which sampling falls below the channel contribution.

\section{Conclusion}
In this research, we introduced a unified embedding framework for
quantum physics-informed neural networks (QPINNs), in which an
embedding is formulated as a functional transformation that shapes the
feature representation presented to the variational quantum circuit.
Within this framework, we proposed two embeddings. LQNN-TE-QPINN
generates data-dependent features with an auxiliary quantum circuit and
combines them linearly with the input coordinates, and AdaFreq-QPINN
promotes the frequency-scaling factors of an arccos--Chebyshev encoding
to trainable parameters. We evaluated both against direct coordinate
encoding and fixed analytical frequency embeddings, together with
alternative hybrid architectures and classical PINN baselines, on one-
and two-dimensional Burgers equations.

At the one-dimensional anchor configuration, LQNN-TE-QPINN attained a
relative $L_2$ error of $0.0916 \pm 0.0021$ with 720 trainable
parameters, compared with $0.4501$ for the direct-encoding QPINN
baseline. It also outperformed LAAF-PINN ($0.1233 \pm 0.0739$) and
PINN ($0.1751 \pm 0.0981$) while requiring approximately eleven times
fewer trainable parameters, and it achieved the lowest $L_\infty$
error among the evaluated models, $0.633 \pm 0.015$. AdaFreq-QPINN
reached $0.3262$, the lowest error among the analytical embeddings at
this configuration, with only four additional trainable parameters
relative to the fixed towers. In two dimensions, LQNN-TE-QPINN attained
the lowest physics-informed training objective at both budgets,
$0.3349$ at 100 epochs and $0.298$ at 200 epochs, and reduced its
solution error from $44.82$ to $40.58$ between the two budgets, which
closed the gap to Reupload-QPINN from $5.93$ to $1.47$.

These results demonstrate that embedding design substantially
influences the approximation and optimization behavior of QPINNs. In
one dimension, the trainable quantum embedding provided clear
improvements over the direct baseline and the fixed analytical
embeddings, and AdaFreq-QPINN indicated that adapting the encoding
frequencies can improve on fixed frequency choices at negligible
parameter cost. In two dimensions, LQNN-TE-QPINN attained the lowest
training objective without attaining a clearly lower solution error
than Reupload-QPINN, which indicates that greater embedding adaptivity
does not by itself guarantee a lower solution error. The same
divergence between the training objective and the solution error
appeared in several one-dimensional comparisons, so the two quantities
should be reported separately when embeddings are compared.

However, the proposed approach still has limitations. First, the
two-dimensional comparison does not establish a final ordering between
LQNN-TE-QPINN and Reupload-QPINN. LQNN-TE-QPINN was still improving at
epoch 200, and the remaining gap is of the same order as the
run-to-run variation we measured for Reupload-QPINN, so the two models
remain unseparated at this configuration. A deterministic re-run with a
longer training budget is the experiment required to resolve this
ordering. Second, LQNN-TE-QPINN retained the lowest absolute derivative
error over the individual noise channels, but it degraded more than
the other models relative to its own noiseless baseline, a behavior
that may stem from noise entering both its embedding and solver
circuits. The noise results therefore describe performance under the
simulated settings considered here and do not constitute evidence of
intrinsic noise robustness. Third, the accessible circuit sizes were
bounded by the memory cost of state-vector simulation with
higher-order automatic differentiation, and the reconstruction
experiment was evaluated on the training grid, so it assesses
reconstruction quality rather than generalization to unseen points.

We will continue to extend the two-dimensional experiments to longer
training budgets under deterministic simulation and to evaluate the
trained models on held-out grids. Concurrently, our research will focus
on shallower trainable embedding circuits and on noise-mitigation
strategies for both the embedding and solver circuits, which the noise
results identify as the main weakness of the proposed model.
Additionally, we intend to apply the embedding framework to broader
classes of PDEs in order to examine whether the observed relationships
between embedding design, optimization behavior, and solution accuracy
persist beyond the Burgers equations.

\section*{Acknowledgment}
The authors acknowledge the support of the Danish e-Infrastructure Consortium (DeiC), the National Quantum Algorithm Academy (NQAA), the Research Center for Systems and Technologies (SYSTEC), the Associate Laboratory Advanced Production and Intelligent Systems (ARISE), the U.S. National Science Foundation (ACCESS) Advanced Computing and Data Resource program, Texas Tech University, and FPT University in Vietnam. The authors used Claude (Anthropic), a large language model, during the preparation of this article. The model was used to verify content and fix grammar mistakes before submitting. All experiments were designed and conducted by the authors, who reviewed, edited, and verified all AI-assisted content and take full responsibility for the content of this article.

\bibliographystyle{IEEEtran}
\bibliography{main}

\phantomsection
\begin{IEEEbiography}[{\includegraphics[width=1in,height=1.25in,clip,keepaspectratio]{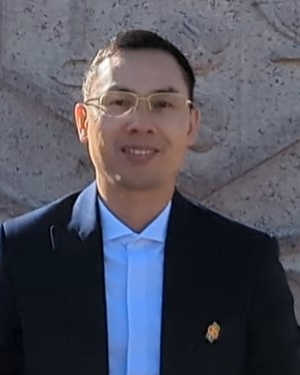}}]{BAN Q. TRAN} received the B.S. degree in Electronics and Communication from Hanoi University of Science and Technology, Vietnam, in 2004 and a dual master’s degree in Computer Science from the University of Science and Technology of Hanoi, Vietnam, and La Rochelle Université, France, in 2021. He is a Ph.D. candidate in Computer Science at Texas Tech University, Texas, USA. From 2017 to 2023, he was a senior lecturer and researcher in Computing Fundamentals at FPT University, Vietnam. His research focuses on new technologies such as Quantum Machine Learning, Quantum Optimization, and Quantum Cybersecurity.
\end{IEEEbiography}

\phantomsection

\begin{IEEEbiography}[{\includegraphics[width=1in,height=1.25in,clip,keepaspectratio]{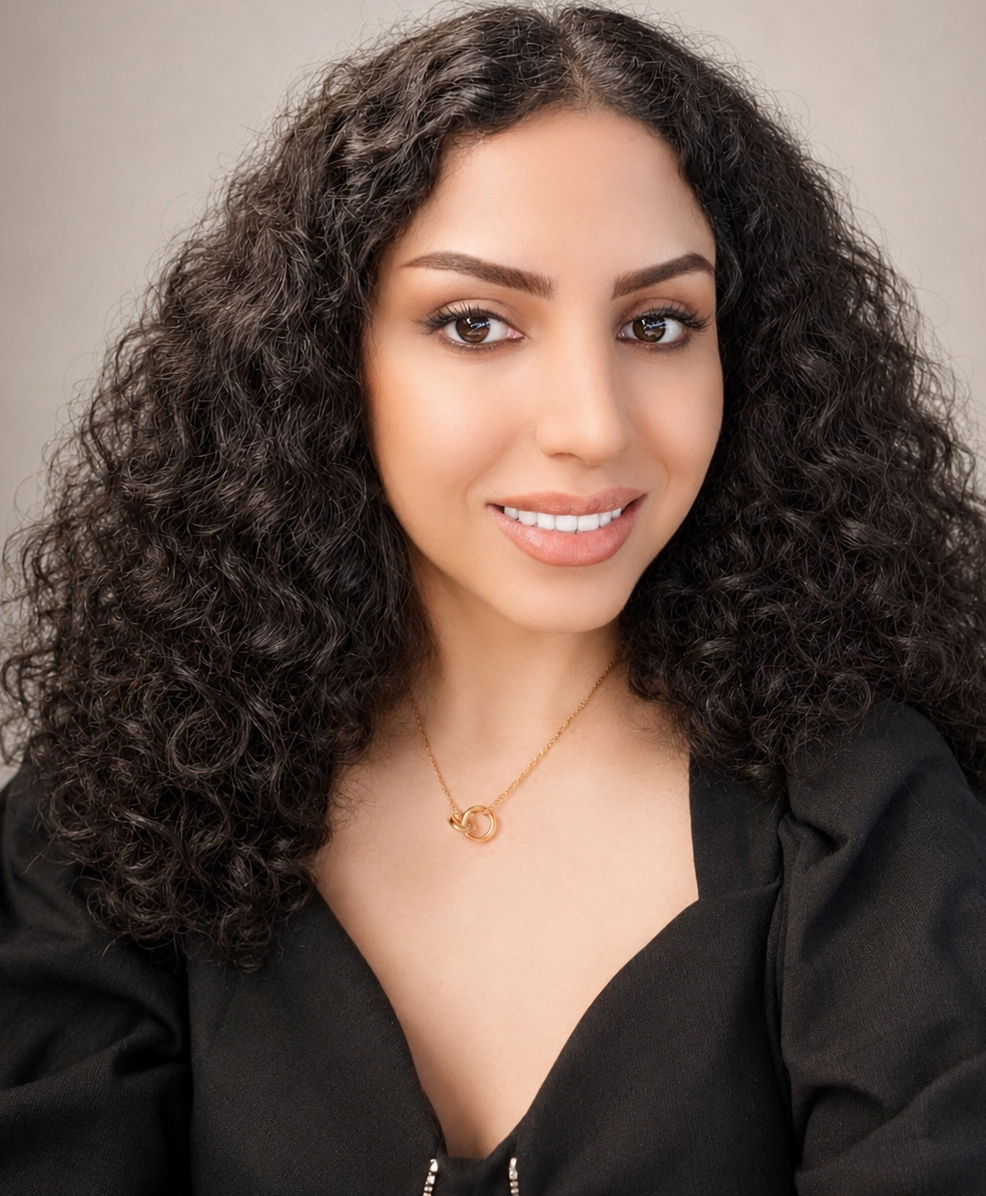}}]{Nahid Binandeh Dehaghani} (Member, IEEE) is a Postdoctoral Researcher in quantum algorithms and quantum software and a member of Denmark’s National Quantum Algorithm Academy. She received the Ph.D. degree in electrical and computer engineering from the University of Porto, Portugal, in 2025, where her research focused on optimal and robust quantum control strategies. Her postdoctoral research focuses on developing innovative quantum-based methodologies and is supported by the Danish e-Infrastructure Consortium (DeiC). Her research expertise spans quantum computing, algorithm design, machine learning, and control, with an emphasis on bridging theoretical insights and practical applications in quantum systems. 
\end{IEEEbiography}

\phantomsection
\begin{IEEEbiography}[{\includegraphics[width=1in,height=1.25in,clip,keepaspectratio]{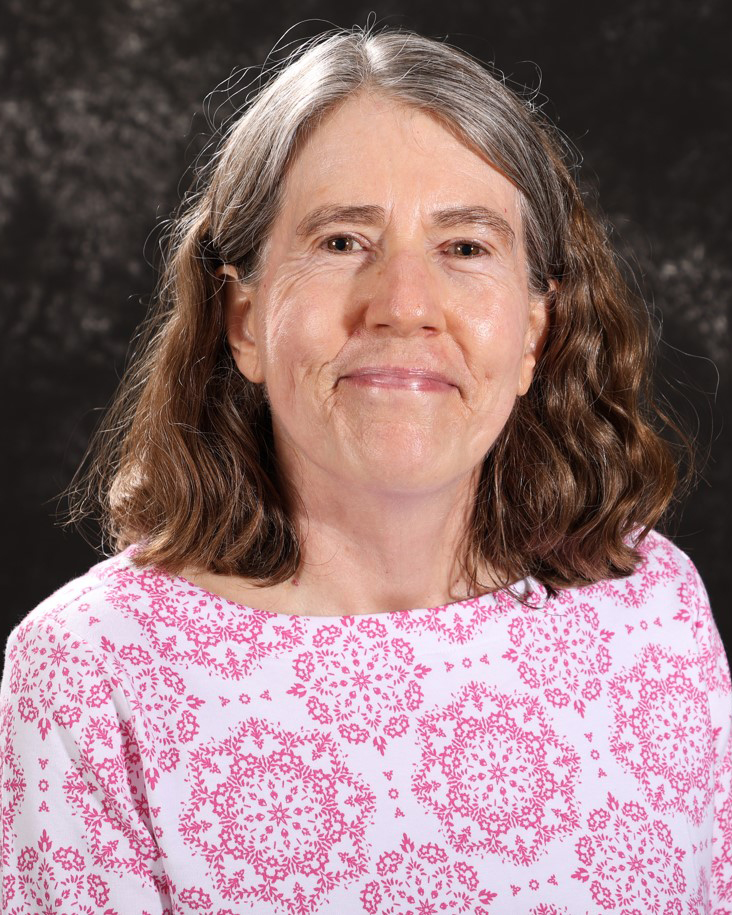}}]{SUSAN MENGEL} received the Ph.D. degree from Texas A$\&$M University in 1990. She is an Associate Professor at Texas Tech University where she is involved in NSF research projects for establishing a variable energy computing cluster, for computing cluster security, and for academic/industry collaborations. She has played strategic leadership roles in numerous trans-disciplinary projects involving the delivery of innovative software and data models in sleep management, student retention and advising, computer education, nutrition, speech therapy, cardiovascular disease, and cybersecurity. She helped to establish the Master's in Software Engineering degree program at Texas Tech University, served as Associate Editor for Computing for the IEEE Transactions on Education, served on the Steering Committee of the ACM/IEEE Computing Curriculum, and served as the Outreach Chair FY19 of the SWE Outreach Committee. She currently serves as Associate Chair on the Texas Tech Institutional Review Board for the Protection of Human Subjects, is the Undergraduate Program Coordinator for the Department of Computer Science at Texas Tech, and is a faculty advisor for the TTU Collegiate Chapter of the Society of Women Engineers.
\end{IEEEbiography}

\phantomsection

\begin{IEEEbiography}[{\includegraphics[width=1in,height=1.25in,clip,keepaspectratio]{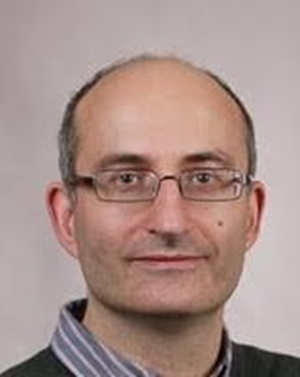}}]{Rafal Wisniewski} (Fellow, IEEE)
earned the Ph.D. degree in electrical engineering in 1997 and the Ph.D. degree in mathematics in 2005, both from Aalborg University, Aalborg, Denmark. He is currently a Professor, the Head of the Learning and Decision Lab, and the Deputy Head for Research at the Department of Electronic Systems, Aalborg University.
His research interests include quantum optimal control, quantum parameter estimation, quantum error correction, quantum machine learning, and quantum optimization. He is actively involved in teaching quantum information and computing and supervises Ph.D., postdoctoral, and master’s students in quantum technologies.
\end{IEEEbiography}

\phantomsection

\begin{IEEEbiography}[{\includegraphics[width=1in,height=1.25in,clip,keepaspectratio]{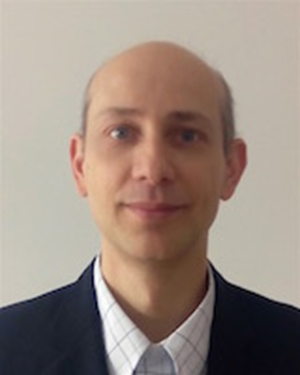}}]{A. Pedro Aguiar} (Senior Member, IEEE) holds a Ph.D. in Electrical and Computer Engineering from IST, University of Lisbon (2002). He is currently a Full Professor at the Faculty of Engineering, University of Porto (FEUP), and serves as Director of ARISE - Advanced Production and Intelligent Systems Associate Laboratory, and Scientific Coordinator of SYSTEC - Research Center for Systems and Technologies. His academic journey includes post-doctoral research at the University of California, Santa Barbara (UCSB) (2002-2005), and teaching positions at FEUP and IST. His research focuses on Control Systems (theory and applications), including motion planning, autonomous robotic vehicles, nonlinear control theory, integration of machine learning with feedback control, quantum feedback systems, and large-scale distributed systems. 
\end{IEEEbiography}

\end{document}